\documentclass[
superscriptaddress,
nofootinbib,
twocolumn,
 amsmath,amssymb,
 aps,
pra,
notitlepage,
longbibliography
]{revtex4-1}

\usepackage{dcolumn}
\usepackage{bm}
\usepackage{epsfig}
\usepackage{graphicx}
\usepackage{latexsym}
\usepackage{amsfonts}
\usepackage{setspace}
\usepackage{graphicx}
\usepackage{amsmath}
\usepackage{mathrsfs}
\usepackage{verbatim}
\usepackage{color}
\usepackage{SIunits}
\usepackage{hyperref}

\usepackage{physics}
\usepackage{soul}

\newcommand{\be}{\begin{equation}}
\newcommand{\ee}{\end{equation}}
\newcommand{\ba}{\begin{array}}
\newcommand{\ea}{\end{array}}
\newcommand{\bqa}{\begin{eqnarray}}
\newcommand{\eqa}{\end{eqnarray}}

\begin{document}

\title{Faithful quantum teleportation via a nanophotonic nonlinear Bell state analyzer}

\author{Joshua Akin} 
\affiliation{Holonyak Micro and Nanotechnology Laboratory and Department of Electrical and Computer Engineering, University of Illinois at Urbana-Champaign, Urbana, IL 61801 USA}
\affiliation{Illinois Quantum Information Science and Technology Center, University of Illinois at Urbana-Champaign, Urbana, IL 61801 USA}
\author{Yunlei Zhao} 
\affiliation{Holonyak Micro and Nanotechnology Laboratory and Department of Electrical and Computer Engineering, University of Illinois at Urbana-Champaign, Urbana, IL 61801 USA}
\affiliation{Illinois Quantum Information Science and Technology Center, University of Illinois at Urbana-Champaign, Urbana, IL 61801 USA}
\author{Paul G. Kwiat} 
\affiliation{Illinois Quantum Information Science and Technology Center, University of Illinois at Urbana-Champaign, Urbana, IL 61801 USA}
\affiliation{Department of Physics, University of Illinois at Urbana-Champaign, Urbana, IL 61801 USA}
\author{Elizabeth A. Goldschmidt} 
\affiliation{Illinois Quantum Information Science and Technology Center, University of Illinois at Urbana-Champaign, Urbana, IL 61801 USA}
\affiliation{Department of Physics, University of Illinois at Urbana-Champaign, Urbana, IL 61801 USA}
\author{Kejie Fang} 
\email{kfang3@illinois.edu}
\affiliation{Holonyak Micro and Nanotechnology Laboratory and Department of Electrical and Computer Engineering, University of Illinois at Urbana-Champaign, Urbana, IL 61801 USA}
\affiliation{Illinois Quantum Information Science and Technology Center, University of Illinois at Urbana-Champaign, Urbana, IL 61801 USA}

\begin{abstract} 

Quantum networking protocols, including quantum teleportation and entanglement swapping, use linear-optical Bell state measurements for heralding the distribution and transfer of quantum information. However, a linear-optical Bell state measurement requires identical photons and is susceptible to errors caused by multiphoton emission, fundamentally limiting the efficiency and fidelity of quantum networking protocols. Here we show a nonlinear Bell state analyzer for time-bin encoded photons based on a nanophotonic cavity with efficient sum-frequency generation to filter multiphoton emissions, and utilize it for faithful quantum teleportation involving spectrally distinct photons with fidelities $\geq 94\%$ down to the single-photon level. Our result demonstrates that nonlinear-optical entangling operations, empowered by our efficient nanophotonics platform, can realize faithful quantum information protocols without requiring identical photons and without the fundamental limit on the efficiency and fidelity of a Bell state measurement imposed by linear optics, which facilitates the realization of practical quantum networks.

\end{abstract}

\maketitle

Quantum information is unique because the linearity of quantum mechanics prevents the cloning of quantum states \cite{wooters1982single}. This is advantageous for information security but a drawback overall, as it prevents the amplification of quantum signals to counteract loss over transmission channels. Quantum teleportation can circumvent this issue by indirectly transferring quantum information using quantum entanglement and a classical communication channel \cite{bennett1993teleporting}, without the original quantum particle needing to traverse the entire distance. The concept can be extended to establish entanglement between distant quantum nodes via entanglement swapping \cite{pan1998experimental}. With a dual network of distributed entanglement and classical communication, it is possible to send quantum information over arbitrary distances \cite{wehner2018quantum,azuma2023quantum}.
Since the first experimental demonstration of quantum teleportation \cite{bouwmeester1997experimental}, tremendous advances have been made toward realizing a practical quantum network \cite{ren2017ground, hu2023progress,liu2024creation,knaut2024entanglement}.

Most quantum networking protocols, including quantum teleportation and entanglement swapping, rely on linear-optical Bell state measurements (LO-BSMs) by interfering two identical photons at a beamsplitter to herald the distribution and transfer of quantum information. However, a LO-BSM requires the two input photons to be identical, leading to false heralds in the presence of multiphoton emission \cite{kok2000postselected,pan2003experimental}, which is ubiquitous among probabilistic quantum light sources and many quantum emitters. This flaw fundamentally limits the efficiency and fidelity of entanglement distribution \cite{sangouard2011quantum,azuma2023quantum}. 
For instance, quantum teleportation using heralded single-photon sources needs substantial attenuation of the source to achieve a high fidelity without  postselection \cite{pan2003experimental}, i.e., without detection of the teleported photon. The latter is necessary for the subsequent manipulation of the teleported photon. Moreover, for entanglement swapping using nonlinear optics-based entangled photon sources and LO-BSMs, the non-postselected fidelity is bounded by $\mathcal{F}\leq 1/3$ (Supplementary Information (SI)). The requirement that the input photons are identical leads to another fundamental constraint on LO-BSMs as any non-identicality reduces the fidelity of the protocol. However, in an actual quantum network with independently produced single and entangled photons traversing dispersive or otherwise unstable channels, it is a major challenge to ensure perfect indistinguishability. Real-world demonstrations of transferring quantum information or distributing entanglement over even a small number of quantum nodes suffer from substantial errors due to this constraint \cite{sun2016quantum,sun2017entanglement,shen2023hertz,knaut2024entanglement}. 

In this work, we demonstrate a nonlinear Bell state analyzer and utilize it for quantum teleportation involving spectrally distinct photons. In contrast to a LO-BSM, a nonlinear Bell state analyzer leverages the sum-frequency generation (SFG) between non-degenerate photons to filter multiphoton emissions from either source, resulting in faithful quantum teleportation and entanglement swapping without the fidelity-efficiency trade-off \cite{sangouard2011faithful}. 
Prior attempts to use SFG for quantum information protocols were limited by the low SFG efficiency of bulk nonlinear crystals and waveguides \cite{kim2001quantum,tanzilli2005photonic,sangouard2011faithful,fisher2021single,guerreiro2013interaction,guerreiro2014nonlinear}, requiring the use of intense optical fields for the SFG. As a result, demonstrating the quantum coherence of the nonlinear interaction at the few- and single-photon levels has proven elusive. 
Our demonstration capitalizes on an In$_{0.5}$Ga$_{0.5}$P nanophotonic cavity with a SFG efficiency transcending the limit of bulk nonlinear crystals and waveguides by several orders of magnitude. With the nanophotonic nonlinear Bell state analyzer, we achieve an average fidelity $\geq 94\%$ for the SFG-heralded quantum teleportation of time-bin encoded photonic states down to the single-photon level and validate the robustness of this scheme against multiphoton emission and over long optical fibers. \\

\begin{figure}[!htb]
\begin{center}
\includegraphics[width=\columnwidth]{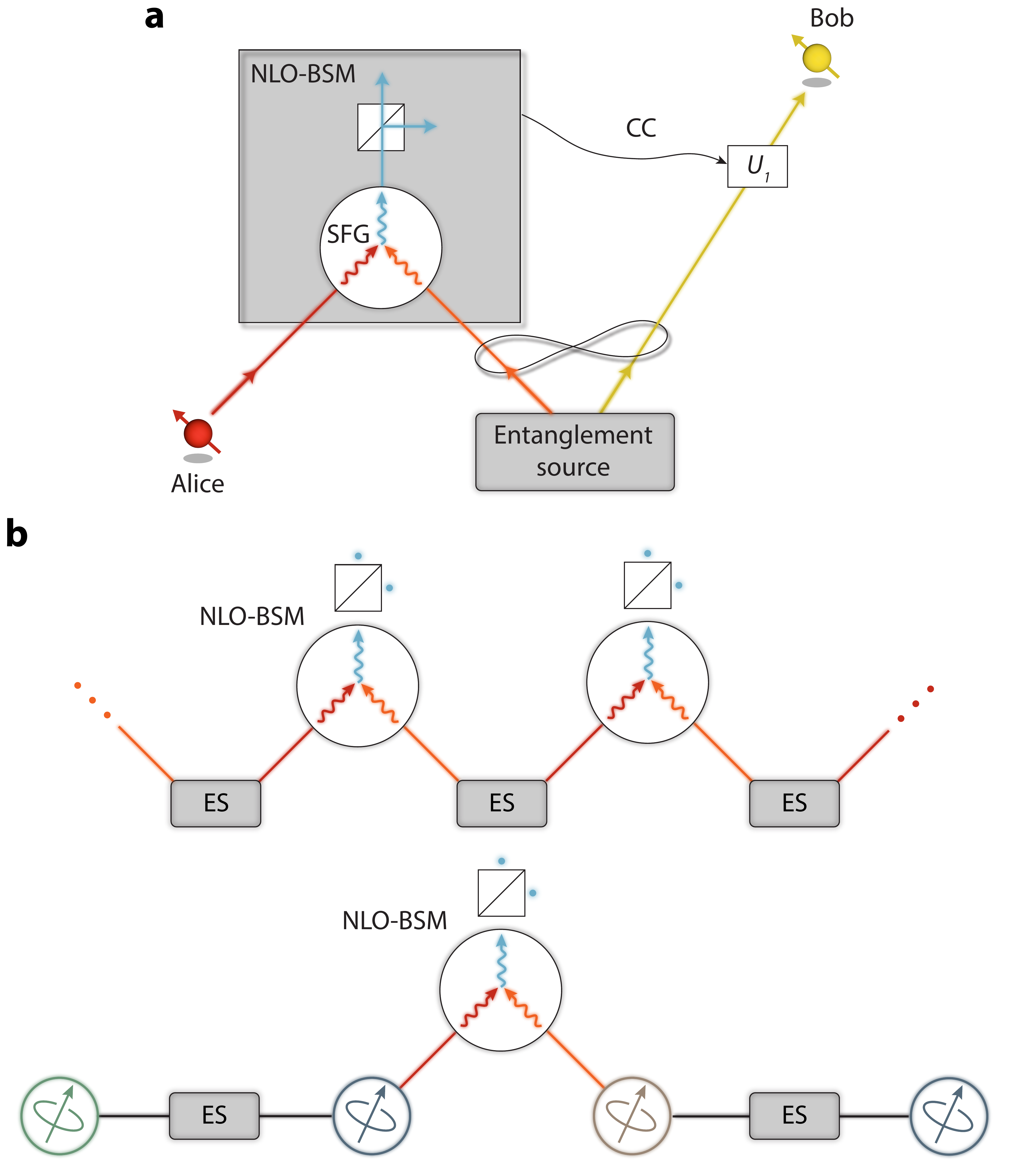}
\caption{\textbf{Nonlinear Bell state analyzer.} \textbf{a}. A schematic of a NLO-BSM that leverages the SFG process being used for quantum teleportation involving spectrally distinct photons. CC: classical communication. \textbf{b}.  Quantum repeater architectures using entanglement swapping between distinct photons (top) and between heterogeneous or distinguishable quantum emitters (bottom) via NLO-BSM. ES: entanglement source.}
\label{fig:scheme}
\end{center}
\end{figure}

For SFG-heralded quantum teleportation (Fig. \ref{fig:scheme}a), a nonlinear optical element supports three-wave mixing between Alice's photon and one photon of the entanglement source to generate a photon at the sum of the two frequencies.  This implements a nonlinear-optical (NLO) BSM when the SFG photon is measured in a basis that encodes the Bell states of the two incoming photonic qubits. The detection of the SFG photon heralds the teleportation of the quantum state of Alice's photon to Bob's photon, up to a single-qubit rotation that depends on the outcome of the measurement of the SFG photon. Because the SFG process is only phase-matched for non-degenerate photons, multiphoton emissions from the same source cannot be up-converted, thus eliminating the vacuum and a portion of uncorrelated photons in the heralded signal. The fidelity of the SFG-heralded teleportation, whether or not it is postselected, is given by (SI):
\begin{equation}\label{eq:SFGFid}
		\mathcal{F} = \left(\frac{1+\sqrt{1-4p_{si}}}{2}\right)^2,
	\end{equation}
where $p_{si}$ is the single-photon-pair probability of the entanglement source. In deriving Eq. \ref{eq:SFGFid}, we assumed thermal number statistics of the photon pair, as is typical for nonlinear optics-based entangled photon sources \cite{braunstein2005quantum}, leading to $p_{si}\leq\frac{1}{4}$, while Alice is an arbitrary pure state. The fidelity exceeds the classical bound of 2/3 even for relatively large $p_{si}(\leq 0.15)$, corresponding to a non-negligible multipair probability, and it is independent of Alice's photon number. This indicates that a true single-photon source is not necessary for the nonlinear Bell state analyzer and sources can be driven more efficiently with less reduction in the fidelity of protocols. The nonlinear Bell state analyzer also avoids the stringent requirement of quantum interference of identical photons, as the two input photons now only need to satisfy the phase-matching condition of SFG---deviating from this condition only affects the efficiency but not the fidelity of the protocol.  
Beyond quantum teleportation, nonlinear Bell state analyzers with even moderate SFG efficiency can enable faithful heralded entanglement swapping \cite{sangouard2011faithful} (SI)---a critical protocol for quantum repeaters (Fig. \ref{fig:scheme}b top). 
Moreover, nonlinear Bell state analyzers can be used to establish entanglement between heterogeneous or distinguishable quantum emitters without the need of wavelength conversion or compensation (Fig. \ref{fig:scheme}b bottom). 

\begin{figure*}[!htb]
\begin{center}
\includegraphics[width=2\columnwidth]{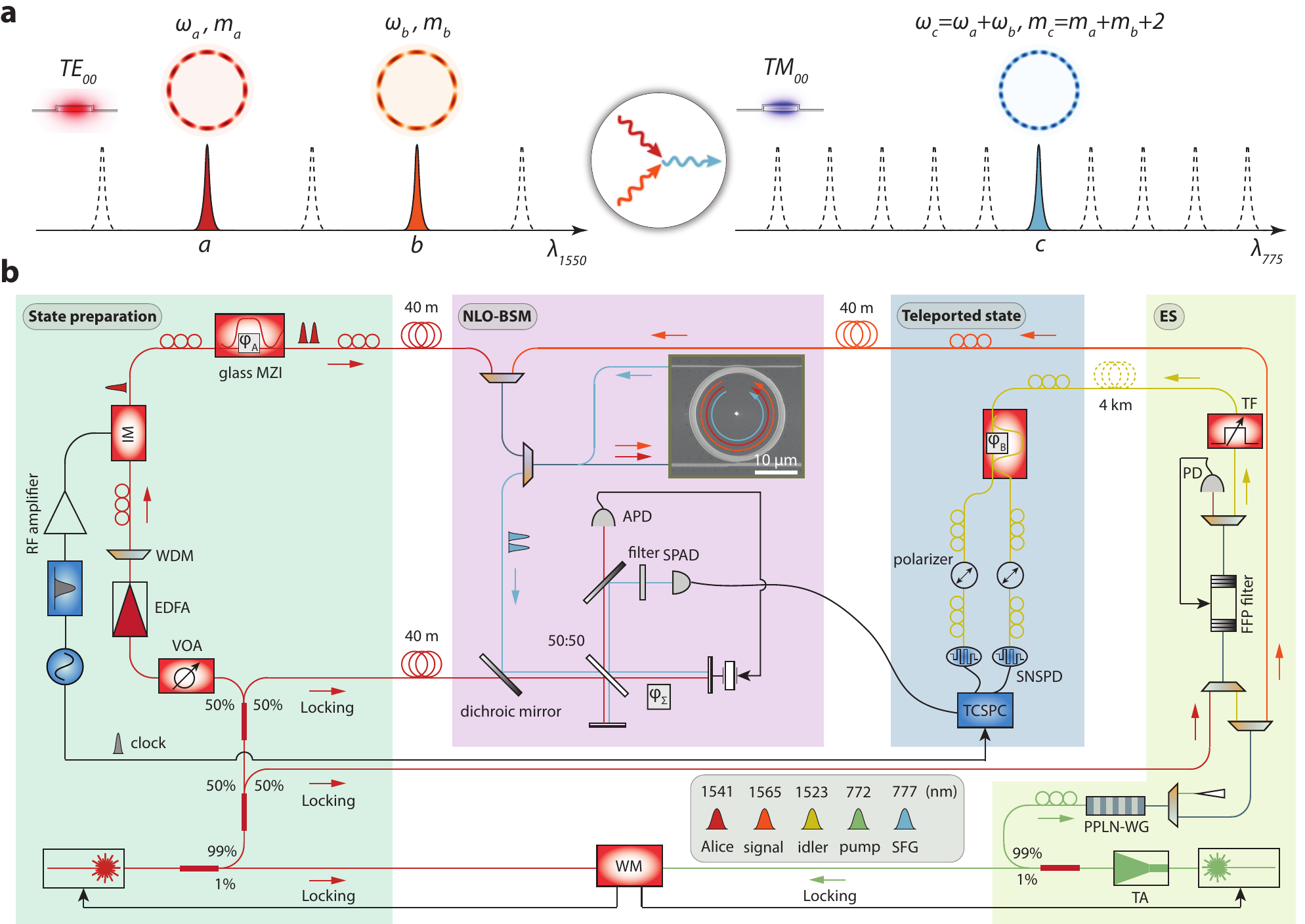}
\caption{\textbf{SFG nanophotonic cavity and quantum teleportation setup.} 
\textbf{a}. Illustration of two TE$_{00}$ and one TM$_{00}$ resonances of an In$_{0.5}$Ga$_{0.5}$P microring resonator satisfying the energy- and phase-matching conditions for the sum-frequency generation process. 
\textbf{b}. Schematic of the experimental setup. A scanning electron microscope image of the In$_{0.5}$Ga$_{0.5}$P nonlinear nanophotonic cavity is shown.  ES: entanglement source. VOA: variable optical attenuator. EDFA: erbium-doped fiber amplifier. WDM: wavelength division-multiplexer. IM: intensity modulator. MZI: Mach-Zehnder interferometer. PPLN-WG: periodically-poled lithium niobate waveguide. FFP filter: fiber Fabry-Perot filter. TF: tunable filter. APD: avalanche photodetector. SPAD: single-photon avalanche diode detector. SNSPD: superconducting nanowire single-photon detector. TCSPC: time-correlated single-photon counting module. WM: wavelength meter. TA: tapered amplifier.}
\label{fig:setup}
\end{center}
\end{figure*}

The nonlinear Bell state analyzer in our experiment is implemented for time-bin encoded photons, which are robust for quantum information transmission in optical fibers. 
Alice's photon is prepared in a superposition state of the early and late time bins, $\ket{\psi_A}=\alpha \ket e_A+\beta \ket l_A$, while the entangled photon source generates a photon pair in the Bell state $\ket{\Phi^+}_{si}=\frac{1}{\sqrt{2}}(\ket e_s\ket e_i+\ket l_s\ket l_i)$. The joint state of the three photons can be expressed as 
\bqa
&&\ket{\psi_A}\otimes\ket{\Phi^+}_{si}\\\nonumber
&=&\frac{1}{2}\big(\ket{\Phi^+}_{As}(\alpha\ket e_i+\beta\ket l_i)+\ket{\Phi^-}_{As}(\alpha\ket e_i-\beta\ket l_i)\\\nonumber
&&+\ket{\Psi^+}_{As}(\alpha\ket l_i+\beta\ket e_i)+\ket{\Psi^+}_{As}(\alpha\ket l_i-\beta\ket e_i) \big),
\eqa
where $\ket{\Phi^\pm}_{As}=\frac{1}{\sqrt{2}}(\ket e_A\ket e_s\pm\ket l_A\ket l_s)$ and  $\ket{\Psi^\pm}_{As}=\frac{1}{\sqrt{2}}(\ket e_A\ket l_s\pm\ket l_A\ket e_s)$ are the four Bell states. SFG can happen in the nonlinear element when both Alice's photon and signal photon are in the same time bin, i.e., in the case of Bell states $\ket{\Phi^\pm}_{As}$. After SFG, the joint state of the initial three photons becomes 
\bqa\label{sfgteleport}
&&\ket{\psi_A}\otimes\ket{\Phi^+}_{si}\xrightarrow{\text{SFG}}\frac{1}{2}\big(\ket{\Sigma^+} (\alpha\ket e_i+\beta e^{-i\varphi_\Sigma}\ket l_i)  \\\nonumber
&&\quad +\ket{\Sigma^-}(\alpha\ket e_i-\beta e^{-i\varphi_\Sigma}\ket l_i) \big),
\eqa
where $\ket{\Sigma^{\pm}}=\frac{1}{\sqrt{2}}( \ket e_\Sigma\pm e^{i\varphi_\Sigma}\ket l_\Sigma)$ are the two orthogonal SFG photon states corresponding to Bell states $\ket{\Phi^\pm}_{As}$ and phase $\varphi_\Sigma$ is introduced during the projection measurement of the SFG photon. A second nonlinear element and delay lines can be introduced to enable SFG between $\ket e$ and $\ket l$, allowing the other two Bell states $\ket{\Psi^\pm}_{As}$ to also be distinguished and thus realizing a complete Bell state analyzer (SI), which is impossible with passive linear optics and unentangled ancillary photons \cite{lutkenhaus1999bell,ewert20143,bayerbach2023bell}. By measuring the SFG photon in the $\ket{\Sigma^{\pm}}$ basis, the idler photon is projected to a single-photon state, which differs from the original Alice's photon up to a single-qubit rotation. 

\begin{figure}[!htb]
\begin{center}
\includegraphics[width=0.8\columnwidth]{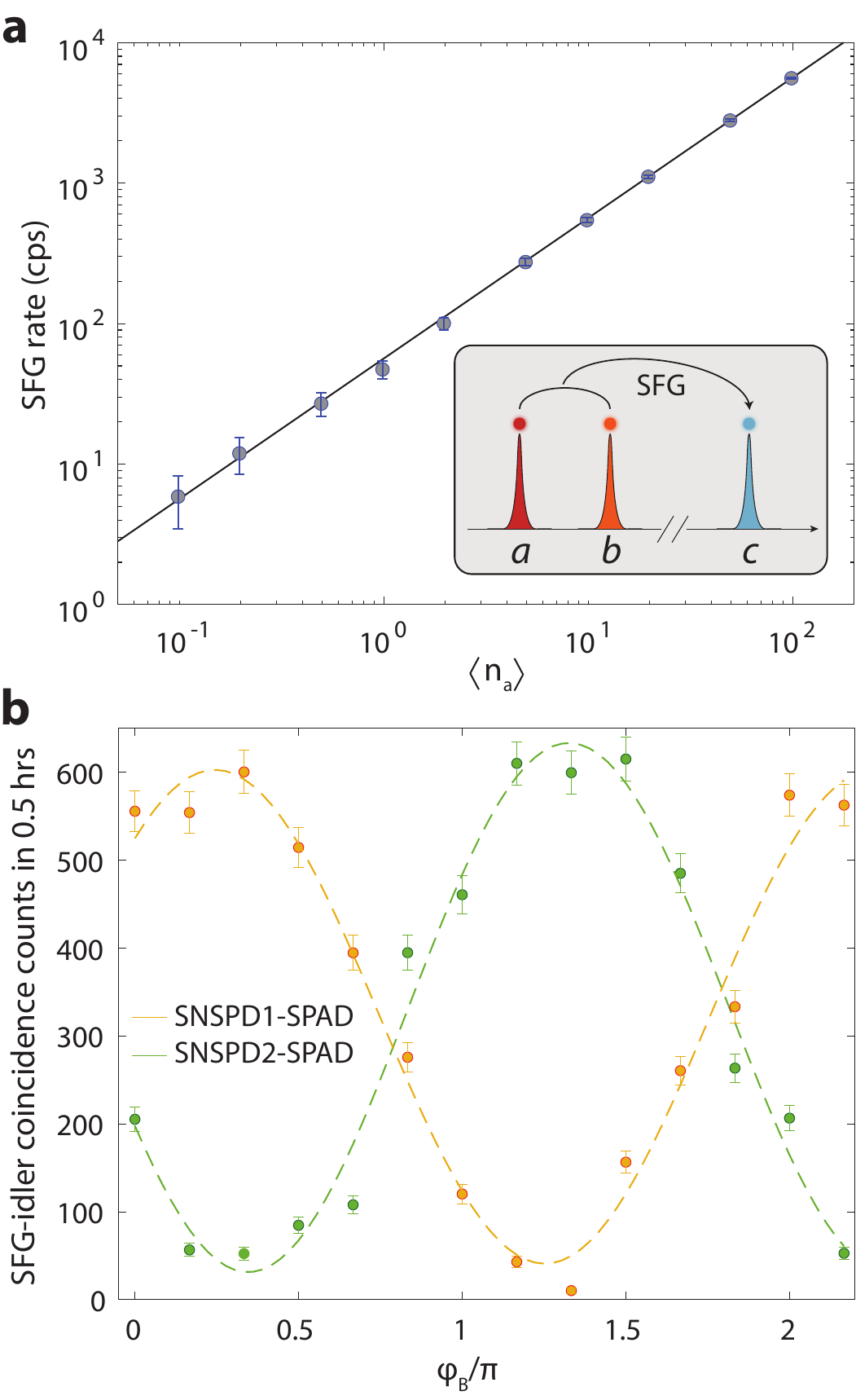}
\caption{\textbf{Entanglement between  time-bin encoded SFG and idler photons.} \textbf{a}. On-chip SFG rate versus cavity photon number $\left<n_a\right>$, with $\left<n_b\right>=5.4\times10^{-3}$.  The line is a linear fit. 
 \textbf{b}. Coincidences between time-bin encoded SFG and idler photons detected by the SPAD and SNSPDs, respectively. SFG cavity photon number $\left<n_{\Sigma}\right>\approx 10^{-5}$. Statistical error bars are one standard deviation. }
\label{fig:characterization}
\end{center}
\end{figure}

Our nonlinear Bell state analyzer exploits a $\chi^{(2)}$ nonlinear microring resonator made from thin-film In$_{0.5}$Ga$_{0.5}$P (Fig. \ref{fig:setup} inset), which possesses a very large second-order nonlinearity, low optical losses, and wavelength-scale optical confinement \cite{zhao2022ingap} to bolster the efficiency of the SFG process.  The In$_{0.5}$Ga$_{0.5}$P microring resonator is phase-matched for two non-degenerate transverse-electric (TE$_{00}$) modes $a$ and $b$ in the telecommunication band and one transverse-magnetic (TM$_{00}$) mode $c$ in the 780 nm band, satisfying the energy- and phase-matching conditions: $\omega_c=\omega_a+\omega_b$ and $m_c=m_a+m_b\pm2$, where $\omega_k$ and $m_k$ are the frequency and azimuthal number of  mode $k$, respectively (Fig. \ref{fig:setup}a and SI). The three-wave-mixing interaction between these modes are described by the Hamiltonian $H/\hbar=g(a^\dagger b^\dagger c+a b c^\dagger)$, where $a^\dagger$($b^\dagger$, $c^\dagger$) and $a$($b$, $c$) are the creation and annihilation operators of mode $a$($b$, $c$) and $g$ is the single-photon nonlinear coupling rate. For the microring resonator used in this experiment, $\lambda_a=1541.010$ nm, $\lambda_b=1565.392$ nm, $\lambda_c=776.553$ nm, and $g/2\pi\approx 14$ MHz.  The linewidth of these resonances is approximately 4 GHz. The microring resonator is coupled with two bus waveguides for transmitting the 1550 nm and 780 nm band light separately and these waveguides are combined at an on-chip wavelength-division multiplexer, allowing the use of a single optical fiber for coupling of all modes with a coupling efficiency of approximately 70\% and 30\% for the 1550 nm and 780 nm light, respectively. The device is mounted in a cryostat (4 K) to prevent wavelength drift of the nanophotonic cavity. We note that with improved engineering of devices for stability the nonlinear Bell state analyzer can operate at room temperature.

The schematic of the experimental setup is depicted in Fig. \ref{fig:setup}. The nonlinear Bell state analyzer is located in one laboratory and the rest of the setup is in another laboratory. Alice's time-bin state is prepared from an attenuated laser pulse, which passes through an unbalanced Mach-Zehnder interferometer (MZI) made from glass integrated photonic circuits to form the early and late time bins with a delay of $\tau=1$ ns. The optical pulse is created by intensity modulation of a continuous-wave laser. Attenuated laser beams can be applied to decoy state protocols developed through quantum cryptography \cite{lo2005decoy} and quantum fingerprinting \cite{arrazola2014quantum}, and are widely used in quantum teleportation experiments \cite{bussieres2014quantum,takesue2015quantum,valivarthi2016quantum,shen2023hertz}. The frequency of Alice's pulse matches mode $a$ of the microring resonator with a pulse width of 330~ps and a repetition rate of 250 MHz. 
The phase difference $\varphi_A$ between the early and late time bins is controlled and stabilized via the temperature of the glass MZI. The microwave pulse that drives the intensity modulator for Alice's pulses serves as the clock for the time-bin pulse sequence. The entangled photon pair is generated via spontaneous parametric down-conversion (SPDC) in a periodically-poled LiNbO$_3$ waveguide pumped by a continuous-wave laser at the pump wavelength $\lambda_p=771.921$ nm. Both Alice's laser and the SPDC pump laser are stabilized by a wavemeter but are not phase-locked to each other. 

The SPDC photon pairs generated by a continuous-wave pump are in a time-frequency entangled state. 
However, only those signal photons that are time-matched with the time-bins defined by Alice's pulses and frequency-matched with mode $b$ of the microring resonator can interact with Alice's photon to generate the SFG photon. Consequently, the frequency- and time-matched signal and idler time-bin qubits are gated by Alice's pulses to form the Bell state $\ket{\Phi^+}_{si}$. The average number of signal photons in the resonator for each signal time-bin qubit is $\left<n_b\right>=5.4\times10^{-3}\ll 1$, and thus there is at most one photon in Alice's time-bin state that is converted to a SFG photon. The idler photon entangled with the signal photon is wavelength-selected by a tunable fiber Fabry-Perot filter with a bandwidth similar to the microring resonator. To herald the quantum teleportation, the SFG time-bin qubit is projected to the $\ket{\Sigma^{\pm}}$ states using a free-space unbalanced Michelson interferometer whose time delay matches that of the time-bin qubit and whose phase delay $\varphi_\Sigma$ is controlled by a motorized stage. Both the fiber Fabry-Perot filter and the Michelson interferometer are stabilized to Alice's laser. Quantum state tomography of Bob's teleported time-bin qubit is performed using a glass integrated photonic circuit MZI with a tunable phase delay $\varphi_B$ to project Bob's qubit to the three axes of the Bloch sphere, from which the density matrix can be constructed. More information about the setup and measurement details is provided in the SI. 

\begin{figure*}[!htb]
\begin{center}
\includegraphics[width=1.7\columnwidth]{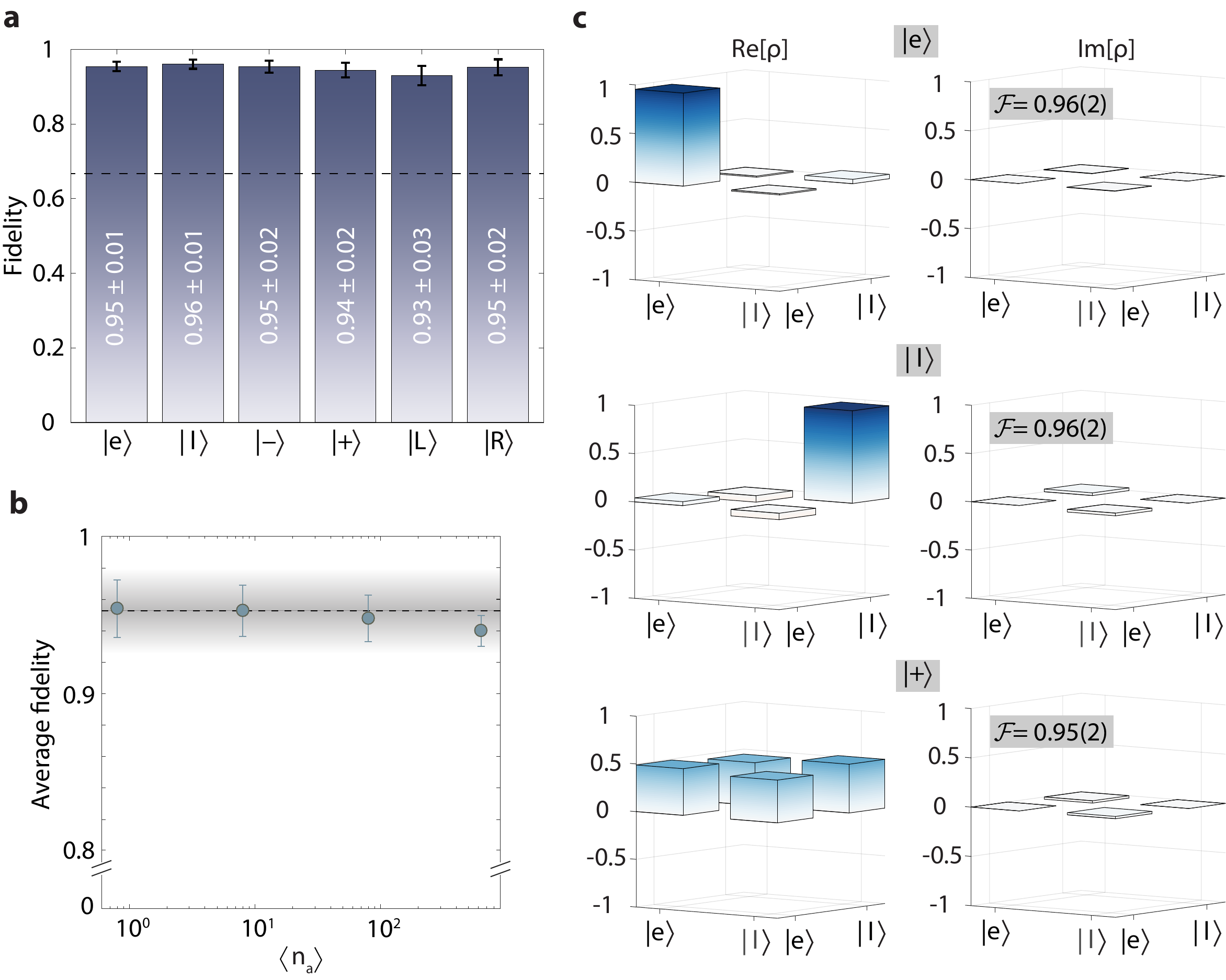}
\caption{\textbf{Experimental results for SFG-heralded quantum teleportation.} \textbf{a}. A summary of the teleportation fidelities for the states $\ket e$, $\ket l$, $\ket -$, $\ket +$, $\ket L$, and $\ket R$. $\left<n_a\right>= 80$. Dashed line indicates the classical bound of $\frac{2}{3}$. Error bars, 1 s.d., deduced from propagated Poissonian counting statistics of the raw detection events and interferometer phase fluctuations. \textbf{b}. Average fidelity of teleported states $\ket e$, $\ket l$, and $\ket +$ for various $\left<n_a\right>$.  Dashed line and shaded area represent the calculated fidelity $\mathcal{F}=(1+\mathcal{V})/2$ and variance $\Delta \mathcal{F}=\Delta \mathcal{V}/2$. \textbf{c}.  Density matrix and fidelity of teleported states $\ket e$, $\ket l$, and $\ket +$ for $\left<n_a\right>=0.8$. }
\label{fig:fidelity}
\end{center}
\end{figure*}

Fig. \ref{fig:characterization}a shows the measured on-chip SFG rate as a function of the mode $a$ cavity photon number, $\left<n_a\right>=P_a/(\hbar\omega_a\kappa_a)$, down to the single-photon level, with the mode $b$ cavity photon number fixed at $\left<n_b\right>=5.4\times10^{-3}$. We inferred a SFG efficiency $\eta_{\mathrm{SFG}}\equiv P_{\mathrm{SFG}}/(P_aP_b)=4.4\times10^4\%$/W, where all the powers are the on-chip power, and the corresponding single-photon SFG probability $p_{\mathrm{SFG}}=4g^2/\bar\kappa_{a,b}\kappa_c\approx 4\times10^{-5}$, where $\kappa_k$ is the dissipation rate of mode $k$ (SI). This marks an improvement over prior nonlinear waveguides by three orders of magnitude  \cite{tanzilli2005photonic,guerreiro2013interaction,guerreiro2014nonlinear,fisher2021single}. 
Next we measured the entanglement between the time-bin encoded SFG and idler photons. The interferometers are stabilized to limit phase variances to $\Delta\varphi_{A, B}=1.1\times10^{-3}\pi$ and $\Delta\varphi_\Sigma=2.4\times10^{-3}\pi$ over one hour (SI). 
After passing through the unbalanced interferometers, the time-bin encoded SFG and idler photons appear in three time bins, where the middle time bin contains a temporally-overlapped superposition of the original early and late states.  From Eq. \ref{sfgteleport}, the coincidence counts between the middle-bin SFG photon $\ket{\Sigma^+}= \frac{1}{\sqrt{2}}(\ket e_\Sigma + e^{i\varphi_\Sigma}\ket l_\Sigma)$ (detected in a single output port with a SPAD) and middle-bin idler photon $\ket{\psi_B}= \frac{1}{\sqrt{2}}(\ket e_i \pm e^{i\varphi_B}\ket l_i)$, where the $\pm$ depends on which output port SNSPD detects the photon, are given by
\begin{equation}
	\begin{gathered}
		C_{\Sigma \vert B} =N_0\left( 1\pm\mathrm{cos}\left(\varphi_A - \varphi_\Sigma - \varphi_B\right)\right),
	\end{gathered}
	\label{eq:Coherence}
\end{equation}
where Alice's photon is in state $\ket{\psi_A}=\frac{1}{\sqrt{2}}(\ket e_A + e^{i\varphi_A}\ket l_A)$. Fig. \ref{fig:characterization}b shows the measured SFG-idler coincidence fringes for the SFG cavity photon number $\left<n_{\Sigma}\right>\approx 10^{-5}$, 
by varying $\varphi_B$ and fixing $\varphi_{A,\Sigma}$. The visibility of the coincidence fringe is $\mathcal{V}=90.5\%$, which exceeds the Clauser-Horne limit of $\frac{1}{\sqrt{2}}\approx70.7\%$ \cite{clauser1974experimental} and verifies the entanglement between the SFG and idler time-bin qubits. The visibility of the coincidence fringe is primarily limited by the imperfect visibility of individual interferometers (SI).

\begin{figure}[!htb]
\begin{center}
\includegraphics[width=0.9\columnwidth]{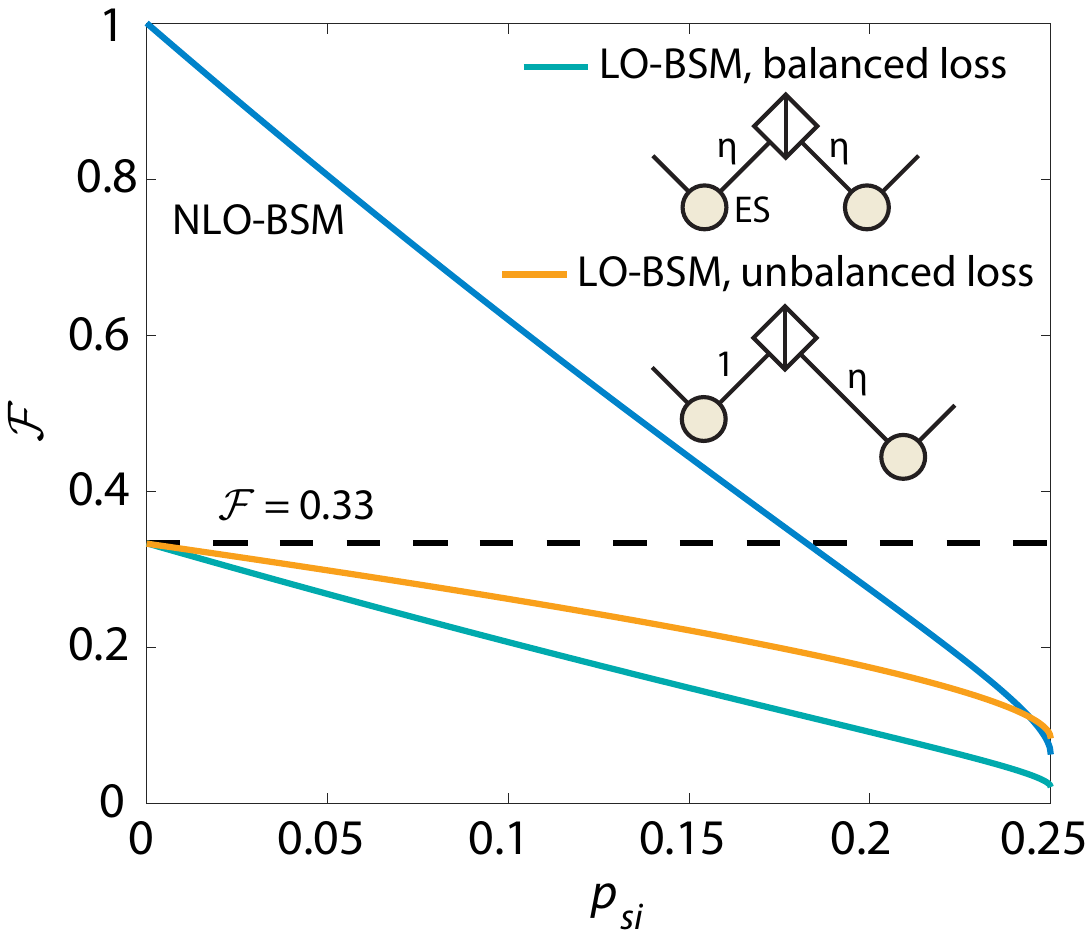}
\caption{\textbf{Optimal fidelity of non-postselected entanglement swapping.}  Green: LO-BSM with balanced loss ($\eta\ll 1$) for two input channels. Yellow: LO-BSM with unbalanced loss. Dashed line indicates the bound $\mathcal{F}=1/3$ for LO-BSM generally. Blue: NLO-BSM, which applies to both cases and the lossless case. Nonlinear optics-based entanglement source (ES) is assumed. For the blue and green curves, $p_{si}$ is the single-pair generation probability of the two ESs. For the yellow curve, $p_{si}$ is the single-pair generation probability of the ES in the lossy channel and that of the ES in the lossless channel is $\eta p_{si}$ for optimal fidelity. See SI.}
\label{fig:eswapping}
\end{center}
\end{figure}

We then used the nonlinear Bell state analyzer for teleportation of different Alice states 
$\ket e$, $\ket l$, $\ket+$, $\ket-$, $\ket R$, and $\ket L$, which are prepared 
by varying $\varphi_A$ or by removing Alice's MZI in the case of $\ket e$ and $\ket l$. The teleportation is heralded by the detection of the SFG photon in the $\ket{\Sigma^+}$ state. The phase of the SFG time-bin qubit is chosen to be $\varphi_\Sigma=0$, such that the teleported state is exactly Alice's state without the need of single-qubit rotation. Quantum state tomography of the teleported state is performed via the time-resolved detection of idler photons conditioned on the detection of $\ket{\Sigma^+}$ state, 
and the density matrix of the teleported state is constructed using maximum likelihood estimation \cite{altepeter2005photonic} (SI). The teleportation fidelity is calculated using the measured density matrix of Bob's photon and the desired state, $\mathcal{F} = \langle \psi_A \vert \rho_B \vert \psi_A \rangle$. The fidelity of the six states for a cavity photon number of Alice $\left<n_a\right>= 80$ is summarized in Fig. \ref{fig:fidelity}a. The average fidelity of the six states, $\bar{\mathcal{F}}=94.4\pm1.8\%$, is well above the classical limit of $\frac{2}{3}$. The uncertainties on the fidelity include contributions from interferometer phase variations and photon shot noise which is simulated using the Monte Carlo method \cite{altepeter2005photonic} (SI). The measured teleportation fidelity is consistent with the theoretical value $\mathcal{F}=(1+\mathcal{V})/2$, indicating the fidelity is limited by the imperfect visibility of interferometers. We also varied the cavity photon number of Alice and performed quantum teleportation of $\ket e$, $\ket l$, $\ket +$ states down to the single-photon level. The average fidelity of the three states versus cavity photon number $\left<n_a\right>$ is shown in Fig. \ref{fig:fidelity}b and the density matrix of the teleported $\ket e$, $\ket l$, $\ket +$ states for $\left<n_a\right>=0.8$ is displayed in Fig. \ref{fig:fidelity}c (see the SI for the density matrix and fidelity of all measured states). The high fidelity ($\geq 94\%$) across all the measured cavity photon numbers proves that the nonlinear Bell state analyzer is immune to multiphoton-driven errors, while, in contrast, a LO-BSM is unfeasible in the high-photon-number regime \cite{valivarthi2016quantum}. 
We also performed SFG-heralded quantum teleportation over long distances using a 4 km-long fiber spool inserted between the entangled photon source and Bob. The fidelity drops slightly to $\bar{\mathcal{F}}=90.5\pm1.6\%$, mainly because the visibility of the glass integrated photonic circuit MZI is polarization sensitive and the polarization drifts in the long fiber during the integration time. 

In summary, we have demonstrated a nonlinear Bell state analyzer and applied it for faithful quantum teleportation of time-bin encoded photonic states down to the single-photon level. Key to this demonstration is a nanophotonic cavity with a single-photon SFG probability ($4\times10^{-5}$) three orders of magnitude greater than prior nonlinear waveguides \cite{tanzilli2005photonic,guerreiro2013interaction,guerreiro2014nonlinear,fisher2021single}.  Using state-of-the-art In$_{0.5}$Ga$_{0.5}$P microring resonators with a nonlinearity-to-loss ratio ($g/\kappa$) of a few percent \cite{zhao2022ingap, akin2024ingap}, we expect to be able to achieve SFG probability ($p_{\mathrm{SFG}}\sim4(g/\kappa)^2$) on the order of $10^{-3}$. Beyond quantum teleportation, the nonlinear Bell state analyzer can enable quantum networking protocols, including heralded entanglement swapping  \cite{sangouard2011faithful, vitullo2018entanglement}, with a much higher fidelity than linear-optical schemes (see Fig. \ref{fig:eswapping} and SI). Given that quantum light sources can be driven more efficiently when using the nonlinear Bell state analyzer, since it directly projects out unwanted multipair noise events, our approach should already achieve a higher entanglement rate than using LO-BSM in some practical scenarios (SI).
Furthermore, teleportation of high-dimensional quantum information via nonlinear interactions is also possible, following the proof-of-principle demonstrations with intense optical fields \cite{sephton2023quantum,qiu2023remote}.\\

\vspace{2mm}
\noindent\textbf{Acknowledgements}\\ 
We thank Liang Jiang and Eric Chitambar for their discussions. This work is supported by US National Science Foundation under Grant No. 2223192 and QLCI-HQAN (Grant No. 2016136) and U.S. Department of Energy Office of Science National Quantum Information Science Research Centers.

\noindent\textbf{Author contributions}\\ 
K.F., E.A.G., P.G.K. proposed the experiment.  J.A. fabricated the device. J.A., Y.Z. performed the experiment and analyzed the data. K.F. supervised the project.  All authors contributed to the writing of the paper.

\end{document}


\title{Supplementary Information for: ``Faithful quantum teleportation via a nanophotonic nonlinear Bell state analyzer'' }

\author{Joshua Akin} 
\affiliation{Holonyak Micro and Nanotechnology Laboratory and Department of Electrical and Computer Engineering, University of Illinois at Urbana-Champaign, Urbana, IL 61801 USA}
\affiliation{Illinois Quantum Information Science and Technology Center, University of Illinois at Urbana-Champaign, Urbana, IL 61801 USA}
\author{Yunlei Zhao} 
\affiliation{Holonyak Micro and Nanotechnology Laboratory and Department of Electrical and Computer Engineering, University of Illinois at Urbana-Champaign, Urbana, IL 61801 USA}
\affiliation{Illinois Quantum Information Science and Technology Center, University of Illinois at Urbana-Champaign, Urbana, IL 61801 USA}
\author{Paul G. Kwiat} 
\affiliation{Illinois Quantum Information Science and Technology Center, University of Illinois at Urbana-Champaign, Urbana, IL 61801 USA}
\affiliation{Department of Physics, University of Illinois at Urbana-Champaign, Urbana, IL 61801 USA}
\author{Elizabeth A. Goldschmidt} 
\affiliation{Illinois Quantum Information Science and Technology Center, University of Illinois at Urbana-Champaign, Urbana, IL 61801 USA}
\affiliation{Department of Physics, University of Illinois at Urbana-Champaign, Urbana, IL 61801 USA}
\author{Kejie Fang} 
\email{kfang3@illinois.edu}
\affiliation{Holonyak Micro and Nanotechnology Laboratory and Department of Electrical and Computer Engineering, University of Illinois at Urbana-Champaign, Urbana, IL 61801 USA}
\affiliation{Illinois Quantum Information Science and Technology Center, University of Illinois at Urbana-Champaign, Urbana, IL 61801 USA}

\maketitle

\tableofcontents

\newpage

\section{Theoretical fidelity}

\subsection{SFG-heralded quantum teleportation}

The normalized state of the SPDC entangled photon source can be written as \cite{braunstein2005quantum}:
\begin{equation}
	\begin{gathered}
		\vert \psi_\mathrm{SPDC} \rangle = \sqrt{1-\epsilon}\sum_{n = 0}^{\infty} \epsilon^{n/2} \vert n \rangle_s\vert n \rangle_i.
	\end{gathered}
	\label{eq:SPDC}
\end{equation}
where $\epsilon$ represents the pump conversion efficiency. Thus the single-photon-pair generation probability is given by $p_{si}=(1-\epsilon)\epsilon\leq\frac{1}{4}$.

Suppose the SFG process involves three modes $a$, $b$, and $c$, and is described by the Hamiltonian $H/\hbar = g\left(abc^\dagger + h.c.\right)$. The evolution of an initial state $\ket{n_an_b0}$ is given by
\begin{equation}
		e^{-iHt/\hbar}\ket{n_an_b0} \approx (1-iHt/\hbar)\ket{n_an_b0} =\ket{n_an_b0}-i \sqrt{p_\mathrm{SFG}n_an_b} \vert (n_a-1) (n_b-1) 1 \rangle).
	\label{eq:SFG}
\end{equation}
Here we have assumed a weak SFG process with $p_\mathrm{SFG}\ll 1$. 

For SFG-heralded quantum teleportation, the frequency of Alice's photon and the signal photon of the entangled photon source matches with that of $a$ and $b$ mode, respectively to generate a SFG photon, while photons from the same source alone do not satisfy the phase-matching condition and cannot generate SFG photons. To calculate the teleportation fidelity, we consider a SPDC source given by Eq. \ref{eq:SPDC} and an arbitrary pure state for Alice, i.e.,
\be
\ket\psi_A=\sum\limits_{m\geq0}c_m\ket m.
\ee
The probabilities of the relevant events are: 
\begin{equation}
		P(\ket\psi_A,n,n) = \Gamma\epsilon^nn,
	\label{eq:linProb2}
\end{equation}
where $\Gamma=p_{\mathrm{SFG}}(1-\epsilon)\sum\limits_{m\geq0}|c_m|^2m$. 

The fidelity of non-postselected teleportation can be written as:
\bqa\label{eq:SFGFid}
		\nonumber\mathcal{F} &=& \frac{P(\ket\psi_A,1,1)}{\sum\limits_{n\geq0}P(\ket\psi_A,n,n)  }\\
		\nonumber&=&(1-\epsilon)^2\\
		&=&\left(\frac{1+\sqrt{1-4p_{si}}}{2}\right)^2.
\eqa
The fidelity is independent of the Alice photon number and, in the weak pump limit, $p_{si}\ll 1$, the fidelity approaches unity. The fidelity exceeds the classical bound of 2/3 even for relatively large $p_{si}(\leq0.15)$.

\subsection{Non-postselected entanglement swapping}

Here we consider entanglement swapping between two SPDC sources described by Eq. \ref{eq:SPDC}. To gain some insight of the LO-BSM approach, first consider only the leading-order effect of two input photons and lossless channels. The fidelity of non-postselected entanglement swapping is given by 
\bqa
		\nonumber\mathcal{F} &=& \frac{P(1_A,1_B)}{P(1_A,1_B)+P(2_A,0_B)+P(0_A,2_B)}\\\nonumber
		&=&\frac{p_{A,si}p_{B,si}}{p_{A,si}p_{B,si}+p_{A,si}^2+p_{B,si}^2}\\
		&\leq& \frac{1}{3},
	\label{eq:swapLOF}
\eqa
where the maximum is achieved when $p_{A,si}=p_{B,si}$. When one of the input channel of the LO-BSM (say from source $B$) has loss $\eta$, the fidelity becomes
\be\label{loeswapest}
\mathcal{F} =\frac{\eta p_{A,si}p_{B,si}}{\eta p_{A,si}p_{B,si}+p_{A,si}^2+\eta ^2p_{B,si}^2}
\ee
and the optimal fidelity can only be achieved by attenuating the lossless source $A$, $p_{A,si}=\eta p_{B,si}$, leading to reduced entanglement rate.

For NLO-BSM, we can calculate the fidelity by accounting all the numbers of photon pairs. The probabilities of relevant events are: 
\begin{equation}
		P(n,m) = (1-\epsilon_A)(1-\epsilon_B)\epsilon_A^{n}\epsilon_B^{m}nmp_{\mathrm{SFG}},
\end{equation}
where $n$ and $m$ are the number of photon pairs emitted by source $A$ and $B$, respectively. The fidelity of entanglement swapping, whether or not it is postselected, is given by
\bqa\label{eq:swapNLOF}
		\nonumber\mathcal{F} &=& \frac{P(1,1)}{\sum\limits_{n,m\geq0}P(n,m) }\\
		\nonumber&=&(1-\epsilon_A)^2(1-\epsilon_B)^2\\
		&=&\left(\frac{1+\sqrt{1-4p_{A,si}}}{2}\right)^2\left(\frac{1+\sqrt{1-4p_{B,si}}}{2}\right)^2.
\eqa

Next we consider practical cases when the input channels of the BSM have losses and the fidelity can be calculated rigorously.

\subsubsection{Balanced loss}

Here we assume the loss of the two input channels of the BSM is $\eta\ll 1$, as in long-distance optical-fiber networks. Because $\eta\ll 1$, we can consider events with only two input photons to the Bell state analyzer, but the source can emit an arbitrary number of photon pairs.

For LO-BSM, the probabilities of relevant events are: 
\begin{equation}
		P(n,m) = (1-\epsilon_A)(1-\epsilon_B)\epsilon_A^{n}\epsilon_B^{m}\left[C_n^2\eta^2(1-\eta)^{n-2}(1-\eta)^{m}+C_n^1C_m^1\eta^2(1-\eta)^{n-1}(1-\eta)^{m-1}+C_m^2\eta^2(1-\eta)^{n}(1-\eta)^{m-2}\right].
\end{equation}
The fidelity of entanglement swapping is given by
\bqa
		\nonumber\mathcal{F} &=& \frac{P(1,1)}{\sum\limits_{n,m\geq0}P(n,m) }\\
		\nonumber&=&\frac{\epsilon_A\epsilon_B}{\frac{\epsilon_A^2}{(1-\epsilon_A(1-\eta))^3}\frac{1}{1-\epsilon_B(1-\eta)}+\frac{\epsilon_A}{(1-\epsilon_A(1-\eta))^2}\frac{\epsilon_B}{(1-\epsilon_B(1-\eta))^2}+\frac{1}{1-\epsilon_A(1-\eta)}\frac{\epsilon_B^2}{(1-\epsilon_B(1-\eta))^3}}\\
		\nonumber&\approx&\frac{1}{3}(1-\epsilon)^4\\
		\nonumber&=&\frac{1}{3}\left(\frac{1+\sqrt{1-4p_{si}}}{2}\right)^4,
\eqa
where the optimal fidelity is achieved for $\epsilon_A=\epsilon_B=\epsilon$.

For NLO-BSM, the probabilities of relevant events are: 
\begin{equation}
		P(n,m) = (1-\epsilon_A)(1-\epsilon_B)\epsilon_A^{n}\epsilon_B^{m}C_n^1C_m^1\eta^2(1-\eta)^{n-1}(1-\eta)^{m-1}p_{\mathrm{SFG}}.
\end{equation}
The fidelity of entanglement swapping is given by
\bqa
		\nonumber\mathcal{F} &=& \frac{P(1,1)}{\sum\limits_{n,m\geq0}P(n,m) }\\
		\nonumber&=&\left(1-\epsilon_A(1-\eta)\right)^2\left(1-\epsilon_B(1-\eta)\right)^2\\
		\nonumber&\approx&(1-\epsilon_A)^2(1-\epsilon_B)^2\\
		&=&\left(\frac{1+\sqrt{1-4p_{A,si}}}{2}\right)^2\left(\frac{1+\sqrt{1-4p_{B,si}}}{2}\right)^2,
\eqa
which is the same as Eq. \ref{eq:swapNLOF}.

\subsubsection{Unbalanced loss}

Here we assume the loss of one input channel (from source $B$) of the Bell state analyzer is $\eta\ll 1$ and that of the other channel (from source $A$) is $\eta=1$. 

For LO-BSM, from the estimation of Eq. \ref{loeswapest}, we know in order to achieve the optimal fidelity when losses are not balanced, source $A$ has to be attenuated to match the received photon flux from source $B$. Because of this, we can consider again events with only two input photons, as probabilities of events with larger photon numbers are small. The probabilities of relevant events are:
\bqa
		 P(0,m) &=& (1-\epsilon_A)(1-\epsilon_B)\epsilon_B^{m}C_m^2\eta^2(1-\eta)^{m-2},\\
		 P(1,m) &=& (1-\epsilon_A)\epsilon_A(1-\epsilon_B)\epsilon_B^{m}C_m^1\eta(1-\eta)^{m-1},\\
		 P(2,m) &=& (1-\epsilon_A)\epsilon_A^2(1-\epsilon_B)\epsilon_B^{m}(1-\eta)^{m}.
\eqa
The fidelity is given by
\bqa
		\nonumber\mathcal{F} &=& \frac{P(1,1)}{\sum\limits_{m\geq0}\left(P(0,m)+P(1,m)+P(2,m)\right) }\\
		\nonumber&=&\frac{\eta\epsilon_A\epsilon_B}{\frac{\eta^2\epsilon_B^2}{(1-\epsilon_B(1-\eta))^3}+\frac{\eta\epsilon_A\epsilon_B}{(1-\epsilon_B(1-\eta))^2}+\frac{\epsilon_A^2}{1-\epsilon_B(1-\eta)}}\\
		\nonumber&\approx&\frac{\eta\epsilon_A\epsilon_B}{\frac{\eta^2\epsilon_B^2}{(1-\epsilon_B)^3}+\frac{\eta\epsilon_A\epsilon_B}{(1-\epsilon_B)^2}+\frac{\epsilon_A^2}{1-\epsilon_B}}\\
		\nonumber &\leq& \frac{1}{3}(1-\epsilon_B)^2\\
		\nonumber&=&\frac{1}{3}\left(\frac{1+\sqrt{1-4p_{B,si}}}{2}\right)^2,
\eqa
where the maximum is achieved when $\epsilon_{A}=\frac{\eta \epsilon_{B}}{1-\epsilon_{B}}$. 

For NLO-BSM, because $\eta\ll 1$, we can consider events with only one input photon from source $B$. The probabilities of relevant events are: 
\begin{equation}
		P(n,m) = (1-\epsilon_A)(1-\epsilon_B)\epsilon_A^{n}\epsilon_B^{m}C_m^1\eta(1-\eta)^{m-1}np_{\mathrm{SFG}}.
\end{equation}
The fidelity of entanglement swapping is given by
\bqa
		\nonumber\mathcal{F} &=& \frac{P(1,1)}{\sum\limits_{n,m\geq0}P(n,m) }\\
		\nonumber&=&\left(1-\epsilon_A\right)^2\left(1-\epsilon_B(1-\eta)\right)^2\\
		\nonumber&\approx&(1-\epsilon_A)^2(1-\epsilon_B)^2\\
		&=&\left(\frac{1+\sqrt{1-4p_{A,si}}}{2}\right)^2\left(\frac{1+\sqrt{1-4p_{B,si}}}{2}\right)^2,
\eqa
which is the same as Eq. \ref{eq:swapNLOF}.

\subsubsection{Rates}

In addition to the fidelity advantage, the NLO-BSM approach can actually be more efficient than the LO-BSM approach in some cases, including the loss-unbalanced case. In particular, the entanglement rate for LO-BSM in this case is given by 
\be
R_{\mathrm{LO}}=p_{A,si}\eta p_{B,si}R_c=\eta^2 p_{B,si}^2R_c,
\ee
where we have used $p_{A,si}=\eta p_{B,si}$ for the optimal fidelity and $R_c$ is the clock rate. In contrast, the entanglement rate for NLO-BSM is given by
\be
R_{\mathrm{NLO}}=p_{\mathrm{SFG}}p_{A,si}\eta p_{B,si}R_c=p_{\mathrm{SFG}}\eta p_{B,si}^2R_c,
\ee
where $p_{A,si}$ can be as large as $p_{B,si}$. When $p_{\mathrm{SFG}}>\eta$, the NLO-BSM approach can be more efficient than the LO-BSM approach. This is relevant, for example, for satellite-mediated entanglement swapping, given the available $p_{\mathrm{SFG}}=10^{-4}-10^{-3}$ using our nanophotonics platform while the loss of satellite-ground link can be $>50$ dB \cite{ren2017ground}.

\section{Cavity SFG efficiency}
For a triply-resonant $\chi^{(2)}$ cavity with phase-matched $a$, $b$, $c$ modes, the Hamiltonian describing the tri-modal interaction is given by
\be
H/\hbar=\omega_a a^\dagger a+\omega_b b^\dagger b+\omega_c c^\dagger c+g(a^\dagger b^\dagger c+abc^\dagger).
\ee
The equations of motion of the three modes are  
\bqa
\frac{da}{dt}=-(i\omega_a+\frac{\kappa_a}{2})a-igb^\dagger c+i\sqrt{\frac{\kappa_{ae}}{2}}a_{\mathrm{in}},\\
\frac{db}{dt}=-(i\omega_b+\frac{\kappa_b}{2})b-iga^\dagger c+i\sqrt{\frac{\kappa_{be}}{2}}b_{\mathrm{in}},\\
\frac{dc}{dt}=-(i\omega_c+\frac{\kappa_c}{2})c-igab+i\sqrt{\frac{\kappa_{ce}}{2}}c_{\mathrm{in}},
\eqa
where $\kappa_k$ and $\kappa_{ke}$ are the total and external dissipation rate of mode $k$, respectively. Here the resonances are coupled bi-directionally to the waveguide, as is the case for the split-resonances of the fabricated microring resonator.

For the SFG process with static driving of modes $a$ and $b$, the static cavity amplitude of the three modes, to the leading order of $g/\kappa$, are solved using the equations of motion:
\bqa
a=-\frac{i\sqrt{\frac{\kappa_{ae}}{2}}}{i(\omega_a-\omega_{pa})+\frac{\kappa_a}{2}}a_{\mathrm{in}},\\
b=-\frac{i\sqrt{\frac{\kappa_{be}}{2}}}{i(\omega_b-\omega_{pb})+\frac{\kappa_b}{2}}b_{\mathrm{in}},\\
c=\frac{igab}{i(\omega_c-\omega_{pa}-\omega_{pb})+\frac{\kappa_c}{2}},
\eqa
where $\omega_{pa}$ and $\omega_{pb}$ are the pump frequencies of mode $a$ and $b$, respectively. 
As a result, we obtain the relation between the SFG power and input pump power:
\be\label{etasfgdef}
P_c=\eta_{\mathrm{SFG}}P_aP_b,
\ee
where $P_c=\frac{\kappa_{ce}}{2}\hbar\omega_c|c|^2$, $P_a=\hbar\omega_a|a_{\mathrm{in}}|^2$, $P_b=\hbar\omega_b|b_{\mathrm{in}}|^2$, with the SFG efficiency given by 
\be
\eta_{\mathrm{SFG}}=g^2\frac{\kappa_{ae}/2}{(\omega_a-\omega_{pa})^2+(\kappa_{a}/2)^2}\frac{\kappa_{be}/2}{(\omega_b-\omega_{pb})^2+(\kappa_{b}/2)^2}\frac{\kappa_{ce}/2}{(\omega_c-\omega_{pa}-\omega_{pb})^2+(\kappa_{c}/2)^2}\frac{\hbar\omega_c}{\hbar\omega_a\hbar\omega_b}.
\ee
In the case with on-resonance driving, $\omega_{pa(b)}=\omega_{a(b)}$, and the frequency-matching condition, $\omega_c=\omega_a+\omega_b$, 
\be\label{etasfg}
\eta_{\mathrm{SFG}}=g^2\frac{\kappa_{ae}/2}{(\kappa_{a}/2)^2}\frac{\kappa_{be}/2}{(\kappa_{b}/2)^2}\frac{\kappa_{ce}/2}{(\kappa_{c}/2)^2}\frac{\hbar\omega_c}{\hbar\omega_a\hbar\omega_b}.
\ee

Next we derive the inherent single-photon SFG probability $p_{\mathrm{SFG}}$ of the nonlinear cavity, which can be defined via
\be\label{psfgdef}
n_c\kappa_c=p_{\mathrm{SFG}}n_an_b\kappa_a.
\ee
The L.H.S. represents SFG photon flux and the R.H.S. represents the flux of $a$ and $b$ photon cluster. We have assumed $\kappa_a\approx \kappa_b$, so the influx photon bandwidth is chosen to be either one of the two modes. Using
\be
n_a=\frac{\kappa_{ae}/2}{(\kappa_a/2)^2}\frac{P_a}{\hbar\omega_a}, \quad
n_b=\frac{\kappa_{be}/2}{(\kappa_b/2)^2}\frac{P_b}{\hbar\omega_b}, \quad
P_c=\hbar\omega_cn_c\frac{\kappa_{ce}}{2},
\ee
and the definitions of Eqs. \ref{etasfgdef} and \ref{psfgdef}, we have
\be\label{relation}
p_{\mathrm{SFG}}=\eta_{\mathrm{SFG}}\frac{(\kappa_{a}/2)^2}{\kappa_{ae}/2}\frac{(\kappa_{b}/2)^2}{\kappa_{be}/2}\frac{\kappa_{c}}{\kappa_a\kappa_{ce}/2}
\frac{\hbar\omega_a\hbar\omega_b}{\hbar\omega_c}.
\ee
Finally, using Eq. \ref{etasfg}, we obtain
\be\label{psfg}
p_{\mathrm{SFG}}=\frac{4g^2}{\kappa_{a}\kappa_{c}}.
\ee
One might use the average value of $\kappa_a$ and $\kappa_b$ in Eq. \ref{psfg}, which makes the equation symmetric for modes $a$ and $b$.
Given the quality factor of the resonances and nonlinear mode coupling $g$, we can estimate $p_{\mathrm{SFG}}$ using Eq. \ref{psfg}. 
For the microring resonator used in the experiment, $g/2\pi\approx 14$ MHz, $Q_{a,b}=4.5\times 10^4$ and $Q_c\approx 10^5$, we find $p_{\mathrm{SFG}}\approx 4\times 10^{-5}$.
In fact, $p_{\mathrm{SFG}}$ can be improved using optimized InGaP microrings. We have realized $Q_{a,b}\approx 4\times 10^5$, $Q_c\approx2\times 10^5$, and $g/2\pi\approx 20$ MHz of 5 $\mu$m rings \cite{akin2024ingap}. With these optimized parameters, $p_{\mathrm{SFG}}\approx  10^{-3}$. We note the non-degenerate nonlinear mode coupling is twice the degenerate nonlinear mode coupling (when $a$ and $b$ modes are the same) \cite{guo2017parametric}, which was measured previously \cite{zhao2022ingap}.

\section{Device}
The SFG device used in the experiment is made from the InGaP nonlinear photonics platform \cite{zhao2022ingap}, which has shown the highest nonlinearity-to-loss ratio among all integrated photonics platforms. This is due to the large nonlinear susceptibility of InGaP ($\chi^{(2)}=220$ pm/V) and its low loss in both 1550-nm and 775-nm wavelength bands. The device pattern is defined using electron beam lithography and transferred to InGaP layer via inductively coupled plasma reactive-ion etch (ICP-RIE) using a mixture of Cl$_2$/CH$_4$/Ar gas. Then a layer of aluminum oxide is deposited via atomic layer deposition. The InGaP device is released from the GaAs substrate using citric acid-based selective etching. More details of the device fabrication can be found in \cite{akin2024ingap}.

The nonlinear cavity used in this experiment is a microring resonator with a radius of 10 $\mu$m. The width of the microring is designed to realize the frequency- and phase-matching condition between two 1550-nm band TE$_{00}$ resonances, $a$ and $b$, and one 780-nm band TM$_{00}$ resonance $c$:
\bqa
\omega_c=\omega_a+\omega_b,\\\nonumber
m_c=m_a+m_b+2,
\eqa
where $m_k$ is the azimuthal number of mode $k$. The $+2$ in the azimuthal number relation is due to the $\chi^{(2)}_{xyz}$ nonlinear susceptibility of InGaP \cite{zhao2022ingap}.  The simulated nonlinear mode coupling between the three modes is $g/2\pi\approx7$ MHz. The microring used in the experiment with a ring width of 670 nm has phase-matched fundamental modes with wavelength $\lambda_{a} = 1541.010$ nm and $\lambda_{b} = 1565.392$ nm and SFG mode with wavelength $\lambda_c = 776.557$ nm. The quality factors of the three resonances are $Q_{a} = 4.4 \times 10^4$, $Q_{b} = 4.7 \times 10^4$, and $Q_{c} \approx 10^5$. The $Q$ of the fundamental resonances is lower than previous result \cite{zhao2022ingap}, because the phase-matching ring width is narrower due to the thinner InGaP film (105 nm) used here. 

The microring resonator is side-coupled to two waveguides to address the 1550 nm and 780 nm band resonances separately. The 1550 nm band TE$_{00}$ resonances are nearly critically coupled with the waveguide, while the 780 nm band TM$_{00}$ resonance is under-coupled due to the more confined electrical field. The bus waveguides are combined in an on-chip wavelength division multiplexer (WDM) with $< 1$ dB wavelength separation/combination loss. The WDM is connected to an adiabatic tapered coupler with a coupling efficiency of about 70\% and 30\% for 1550 nm TE and 775 nm TM polarized light, respectively. This construction enables coupling with only one tapered optical fiber for both input and output. After characterization of the system efficiency, we measured an on-chip SFG efficiency of $\eta_{\mathrm{SFG}} = 4.4\times 10^4 \%/$W for a phase-matched microring.

Phase-matching condition is satisfied in practice by sweeping the ring width in an array of microrings. 
The microring resonances have a small anharmonicity leading to second-harmonic generation (SHG) and SFG being phase matched at slightly different ring widths. The SHG phase matching microring is found by scanning the fundamental frequency laser and measuring the second-harmonic signal. Then to find SFG microring we reverse the nonlinear process and pump microrings around the SHG microring at the second-harmonic frequency. We sort the generated SPDC photons into degenerate and nondegenerate frequencies using a CWDM. This determines both the strongest SFG devices, as well as the mode order of the signal and idler. The final SFG device is then the one with highest efficiency and that satisfies wavelength constraints of the overall experimental setup discussed in Section \ref{sec:lambda}.

\section{Setup and measurement details}
\subsection{Measurement details}
The teleportation setup is illustrated in the main text and here we provide more information about the measurement process and setup. Alice starts from a continuous-wave (CW) diode laser set to 1541.010 nm, which is wavelength-locked using active feedback from a wavemeter. Fractions of the CW light are picked off by beamsplitters to lock other phase-sensitive components, including the free-space Michelson interferometer used for SFG photon interference and the tunable fiber Fabry-Perot (FFP) filter used to spectrally select Idler photons. Then the CW laser power is adjusted with a variable optical attenuator (VOA) and an erbium-doped fiber amplifier (EDFA). Amplified spontaneous emission (ASE) noise of the EDFA is filtered using coarse wavelength-division multiplexers (WDMs). For single-photon level pulses for Alice, the EDFA is removed. 
Alice is then intensity-modulated into a sequence of Gaussian pulses with a repetition rate of 250 MHz and a pulse width of 330 ps. This can constitute as an early time-bin state as is, or a late time-bin after adding an optical fiber equivalent to a delay of 1.05 ns. To generate other states on the Bloch sphere, we used a glass Mach-Zehnder interferometer (MZI) that produces an equal superposition of early and late time-bins with a controllable relative phase $\varphi_A$. The phase is controlled through a heating element and the temperature of the glass MZI is monitored and actively stabilized with a thermistor and PID loop.

The entangled photon pair is generated via spontaneous parametric down-conversion (SPDC) through a magnesium-doped periodically-poled lithium niobate (Mg:PPLN) waveguide. The pump is from a continuous-wave diode laser set to 771.921 nm which is wavelength-locked through a wavemeter. The pump is amplified up to 300 mW using a tapered amplifier (TA). Residual pump light after the PPLN waveguide is filtered using a 1550-nm/780-nm WDM. The SPDC light has an overall power of 80 nW and a 100-nm bandwidth centered at twice the pump wavelength. The signal and idler photon pair is picked out first by a coarse WDM with a channel bandwidth of 20 nm. The signal photons are combined with Alice in another coarse WDM before sent to the SFG device.

In another lab, the combined signal and Alice photons pass through a 1550 nm/780 nm WDM to get to the device. The 1550 nm/780 nm WDM is needed to separate the SFG photons and reflected 1550 nm band photons from the device. The nonlinear cavity is placed inside a cryostat to prevent the drift of resonance wavelengths, which occurs in an ambient environment. Alternatively, one could use active temperature control to compensate the wavelength drift of the cavity in air. A customized microscope system is used for imaging and alignment of the tapered fiber and device in the cryostat. The tapered fiber couples both 1550 nm and 780 nm band light from a single-mode telecom fiber to the on-chip device and vice versa. The signal photon that undergoes SFG is selected out of the broadband SPDC light by the cavity resonance at 1565.392 nm. The signal photon and Alice generate SFG photons which take the reverse path to exit the device. Once off the chip, the 1550-nm/780-nm WDM directs the SFG photon towards the free-space Michelson interferometer, where the SFG photons are projected into the states $\ket{\Sigma^\pm}$. The phase of the Michelson interferometer, $\varphi_\Sigma$, is controlled via a piezo stage and monitored using fractions of the wavelength-locked Alice laser that are also fed into the Michelson interferometer. Dichroic mirrors are used to combine and separate the Alice laser light and SFG photons in the interferometer. After interference, the SFG light is directed to a single-photon avalanche detector (SPAD) with a dark count rate of 60 Hz. 

While the SFG photon is generated and detected, idler photons need to be analyzed to evaluate if teleportation occurs. After separation from the signal photon using a coarse WDM, idler photons are further filtered to a similar bandwidth as the cavity resonance using a tunable FFP filter (FSR 9.1 nm (1.1 THz), bandwidth 2.3 GHz). The FFP filter is stabilized using a fraction of the wavelength-locked Alice laser. All the resonances of the FFP filter are intrinsically locked together as they are determined by the same cavity, allowing using wavelength-locked Alice to stabilize the idler wavelength. After the FFP filter, the Alice laser light is separated from idler using a coarse WDM and idler continues to a tunable bandpass filter with a bandwidth of about 1 nm. The purpose of this filter is to filter light from a few other resonances of the FFP filter that lie in a coarse WDM channel. After the filtering, the idler photon has a wavelength of 1522.870 nm, which is entangled with the cavity-filtered signal photon, as they satisfy $1/\lambda_s+1/\lambda_i=1/\lambda_p$. The idler photon is analyzed in a glass MZI, which projects idler to the six poles of the Bloch sphere. The phase of the glass MZI, $\varphi_B$, is controlled through a heating element and the temperature of the MZI is monitored and actively stabilized with a thermistor and PID loop. Because of fabrication imperfection, the glass MZI can change and mix polarizations as light passes through it. Thus, to purify the state, polarization controllers and polarizers are used to balance the time bins before being detected at the superconducting nanowire single photon detector (SNSPD) with a dark count rate of 100 Hz.

For the teleportation measurement, we set $\varphi_\Sigma = 0$, resulting in Bob receiving Alice's state directly.  To set the phase of the sum-frequency interferometer, we send in continuous-wave light locked to the wavelength of the generate SFG. Actuating the piezo stage allows tracing out an interference fringe for the coherent light with the SFG frequency (note the SFG light generated from the cavity cannot interfere because its coherence time, i.e., roughly the inverse of the cavity linewidth, is shorter than the delay of the interferometer). 
The locking laser, which is a fraction of Alice's laser, also passes through the interferometer observing interference fringe.  
The phases of the two interference fringes are fixed to each other, allowing monitoring of only the locking laser transmission to set and lock the sum-frequency phase.  Bob can project the teleportation output into each pole of the Bloch sphere to reconstruct the density matrix of the teleported state. The output of Bob's MZI contains three time bins. The population of the early and late bins is directly proportional to the early and late state weight in the input state.  Meanwhile the middle time-bin of the MZI output contains a mixture of early and late input photons with the phase $\varphi_B$. By setting $\varphi_B = 0$, the two MZI output middle bins are projected into the $\vert \pm \rangle$ states (note the two outputs of the MZI are $\pi$ out-of-phase). Instead, by setting $\phi_B = \pi/2$, the two middle bins are projected into the $\vert L \rangle$ and $\vert R \rangle$ states. Bob's phase $\varphi_B $ can be set by passing through a continuous-wave 1550-nm band laser locked to the idler frequency before teleportation. Then the temperature of the MZI can be tuned and the transmission of the laser indicates the phase between the arms of the MZI. This must be tuned and set before teleportation since the idler light will not on its own display interference given the short coherence time.

Across the setup there are multiple monitor points to ensure faithful operation of the system over long collection periods. Alice's polarization is monitored through the second port of the MZI where Alice's qubit state is defined. Any device drift or signal/Alice polarization changes into the device are monitored with the 1550-nm band reflection off the device. Finally, the SNSPDs collecting idler counts can be monitored for any variations or drifts in the idler path.

\subsection{Wavelength conditions}\label{sec:lambda}

Our SFG-heralded teleportation experiment involves multiple wavelengths, corresponding to Alice ($\lambda_A$), SFG ($\lambda_\Sigma$), signal ($\lambda_s$), idler ($\lambda_i$), and SPDC pump ($\lambda_p$). The triply-resonant nonlinear cavity supports resonances of wavelength $\lambda_a$, $\lambda_b$, and $\lambda_c$. They need to satisfy a group of conditions to ensure the entanglement of the SFG and idler photons.

(a) Alice and SFG wavelengths match with $\lambda_a$ and $\lambda_c$, respectively, which determines signal wavelength $1/\lambda_s=1/\lambda_\Sigma-1/\lambda_A\approx 1/\lambda_b$. In addition, $\lambda_b-\lambda_a=n\times\lambda_{\mathrm{FSR}}$, where $\lambda_{\mathrm{FSR}}$ about 11.9 nm is the free-spectral range (FSR) of the 1550 nm band TE$_{00}$ resonances of the $R=10\ \mu$m microring. Often not both $\lambda_a$ and $\lambda_b$ are within the wavelength range of the tunable telecom laser ($1520-1570$ nm), so direct SFG characterization of the device is impossible with two telecom lasers. In such a case, we can characterize the device using non-degenerate SPDC or difference frequency generation (DFG) if the SFG wavelength is within the 780-nm band laser range ($765-781$ nm). 

(b) The idler is picked out by a resonance of the FFP filter, which is stabilized by the wavelength-locked Alice laser. As a result, the idler and Alice must be separated by an integer number of FFP filter FSRs, $\lambda_A-\lambda_i=m\times\lambda_{\mathrm{FFP,FSR}}$, where $\lambda_{\mathrm{FFP,FSR}}=9.1$ nm.

(c) The wavelengths of the signal and idler determines the SPDC pump wavelength due to energy conservation, i.e., $1/\lambda_p=1/\lambda_s+1/\lambda_i$. The commercial Mg:PPLN waveguide has a pump wavelength range of 770.5 nm to 773.5 nm, corresponding to a temperature range of $23-75$ $^\circ$C.

(d) Since the wavelength-locked Alice laser is used to stabilize phase-sensitive components in the setup, including the FFP filter and the free-space Michaelson interferometer, this defines another constraint on Alice, signal, and idler wavelengths: they must fall in separate coarse WDM channels to allow efficient separation and combination of these telecom-band wavelengths across the setup. The coarse WDM channels have a bandwidth of about 20 nm.

\begin{figure}[!htb]
	\begin{center}
		\includegraphics[width=0.5\columnwidth]{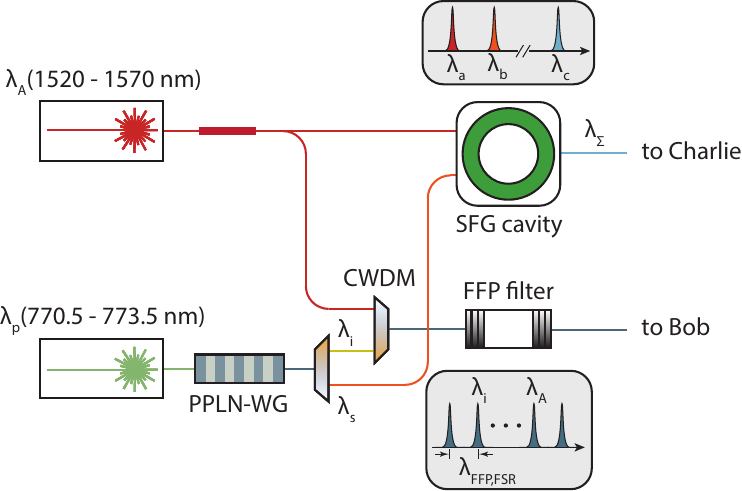}
		\caption{\textbf{Wavelength conditions.} Illustration of the wavelength condition set by various components in the experimental setup.}
		\label{fig:setup}
	\end{center}
\end{figure}

The conditions on the wavelengths and relevant components responsible for the wavelength condition are illustrated in Fig. \ref{fig:setup}.

\section{Conditions of the quantum teleportation}
\subsection{Teleportation distance}
\begin{figure*}[!htb]
	\begin{center}
		\includegraphics[width=0.5\columnwidth]{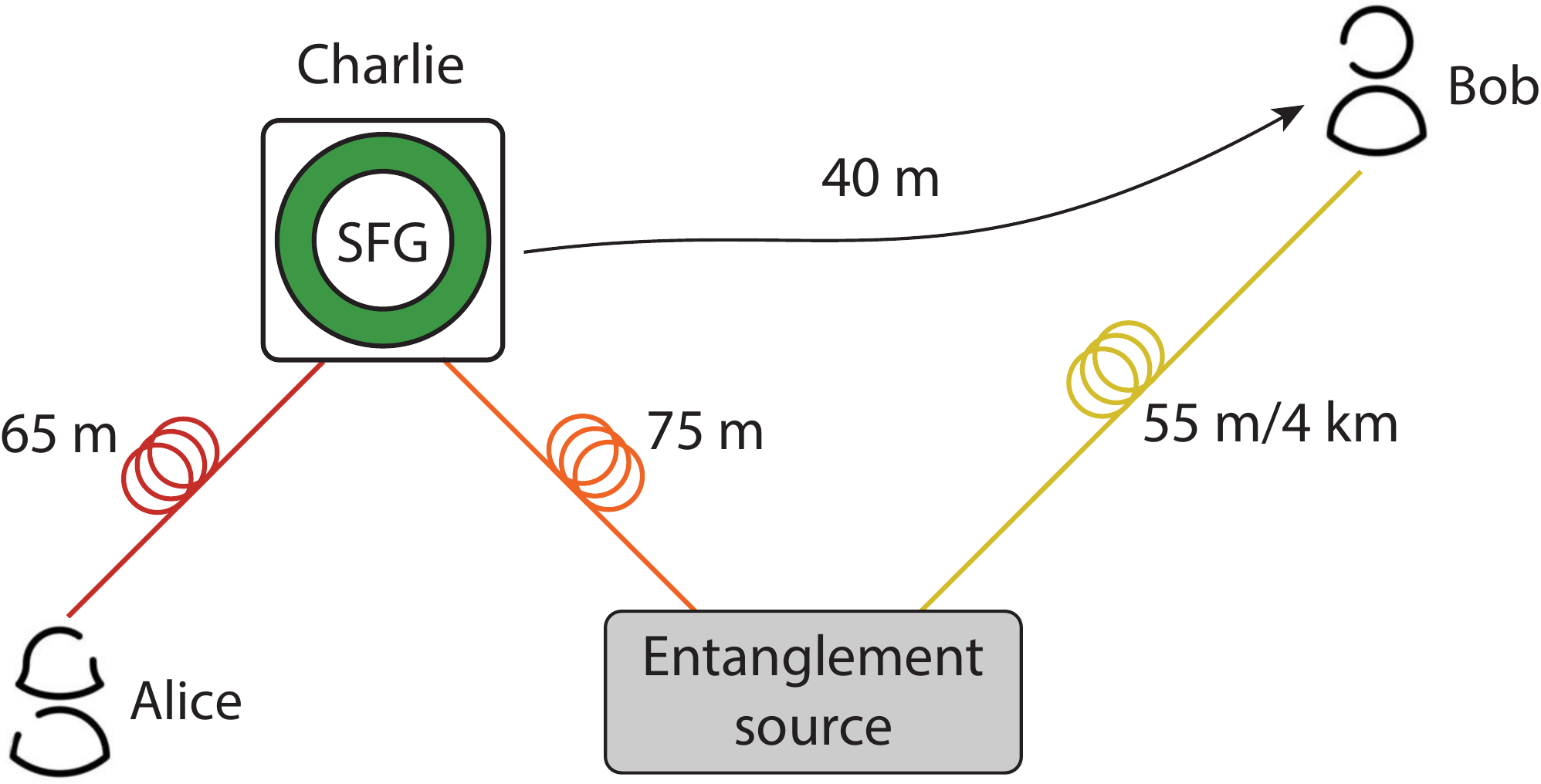}
		\caption{\textbf{Teleportation Distance.} Illustration of the distance between various parties involved in the quantum teleportation.}
		\label{fig:distance}
	\end{center}
\end{figure*}

The physical distances, i.e., optical fiber and cable lengths, between major components involved in the quantum teleportation are illustrated in Fig. \ref{fig:distance}. Alice, Bob, and the entanglement source are located in Lab A, while the SFG-BSM is performed in a cryostat in Lab B. The optical fiber length between Bob and the entanglement source can be made long by insertion of a 4-km-long fiber spool.

\subsection{Interferometer visibility}
\begin{figure*}[!htb]
	\begin{center}
		\includegraphics[width=0.9\columnwidth]{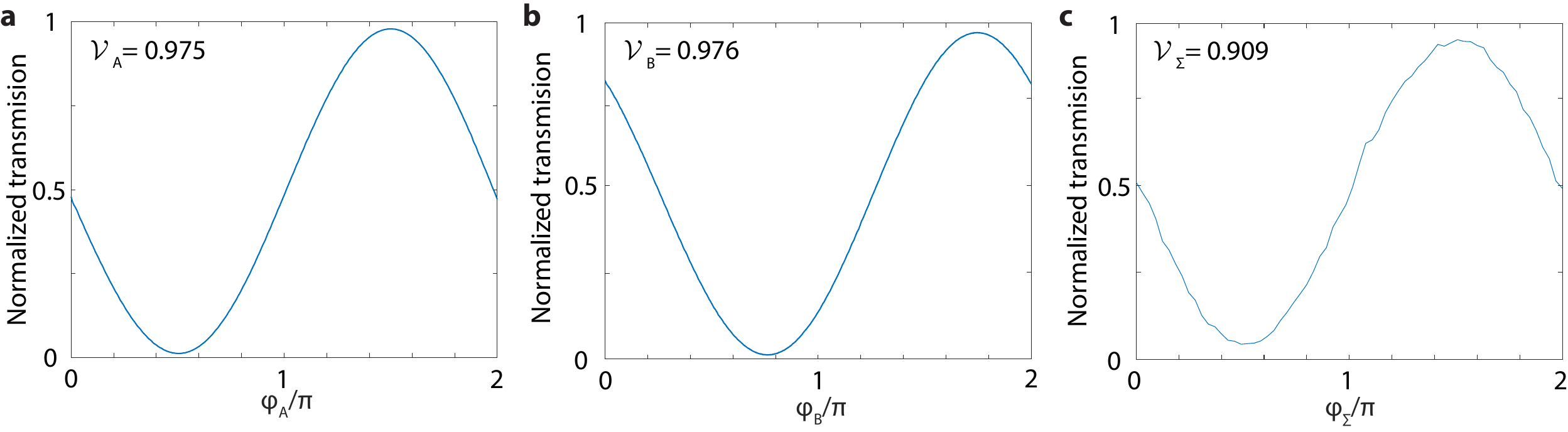}
		\caption{\textbf{Interferometer visibility.} \textbf{a-c} Normalized interference fringe and fringe visibility of the Alice (\textbf{a}), idler (\textbf{b}), and SFG (\textbf{c}) interferometers. }
		\label{fig:visibility}
	\end{center}
\end{figure*}

To achieve a high fidelity of the quantum teleportation, the individual interferometers have to maintain a high visibility. Alice and Bob's interferometers are made from glass integrated photonic circuits, and thus the visibility is determined by the fabrication and cannot be adjusted. The interference fringes of these two interferometers at the Alice and idler wavelengths are show in Fig. \ref{fig:visibility}a and b, respectively. The visibility is $\mathcal{V}_A = 97.5\%$ and $\mathcal{V}_B = 97.6\%$, respectively. The SFG interferometer is a free-space Michelson interferometer. The interference fringe of a continuous-wave coherent light at the SFG wavelength is shown in Fig. \ref{fig:visibility}c with a visibility of $\mathcal{V}_\Sigma = 90.9\%$.

\subsection{System efficiency}
The transmission and detection efficiency of the experimental setup is summarized in Table \ref{tab:loss}. Alice's transmission $t_A$ is measured from the Alice interferometer output to the nonlinear cavity. Signal transmission $t_s$ is measured from the PPLN waveguide output to the nonlinear cavity. Idler transmission $t_i$ is measured from the PPLN waveguide output to the SNSPDs. The SFG transmission $t_\Sigma$ is measured from the nonlinear cavity to the SPAD.  SPDC rate $R_{si}$ out of the PPLN waveguide is 50 MHz in 1 GHz bandwidth.

{\renewcommand{\arraystretch}{1.3}
\begin{table}[h!]
\centering
\caption{\textbf{Summary of system efficiency}}
\vspace{5pt}
\begin{tabularx}{0.8\textwidth} {m{10cm}  m{2.8cm} c m{2cm} }
    \hline
    \hline
       Alice transmission efficiency & $t_A$ &  0.28  \\
       signal transmission efficiency  & $t_s$ &  0.19  \\
       idler transmission efficiency  &  $t_i$  & 0.02 \\ 
       SFG transmission efficiency   & $t_\Sigma$   &  0.08 \\
       SPDC pair rate (in 1 GHz bandwidth) &  $R_{si}$  &  50 MHz \\
	idler detector efficiency &  $\eta_i$  & 0.90  \\
	SFG detector efficiency &  $\eta_\Sigma$  & 0.65 \\
    \hline
    \hline
    \end{tabularx}
    \label{tab:loss}
\end{table}

\subsection{System stability}

The system is highly stable on the order of a few of hours. The locking of wavelengths, the FFP filter locked by the Alice's laser, and any polarization across the setup are in general stable for days on end. The main limiting factor comes from the three interferometers. The free-space interferometer for the SFG signal drifts extremely slowly and can be stably locked with a feedback loop for around a day. The glass PLC MZI for Alice and Bob are the main weaknesses for stability stemming from the temperature instability of the glass PLC units. A PID loop actively fixes the temperature according to a small thermistor centered on the PLC unit. The PLC units are enclosed in plastic enclosures and further wrapped in insulating foam to limit the influence of external temperature changes. However, the phase still varies around $3 \times 10^{-3} \pi$ for a few hours. The stability over a one-hour window is depicted in Fig. \ref{fig:stability}. This is an acceptable amount of drift resulting in $< 0.2 \%$ error on the teleportation fidelity and is much less than the contribution from shot noise. Thus for few- and single-photon data collected over tens of hours the phases are checked and adjusted as needed every few hours.

\begin{figure*}[!htb]
	\begin{center}
		\includegraphics[width=0.5\columnwidth]{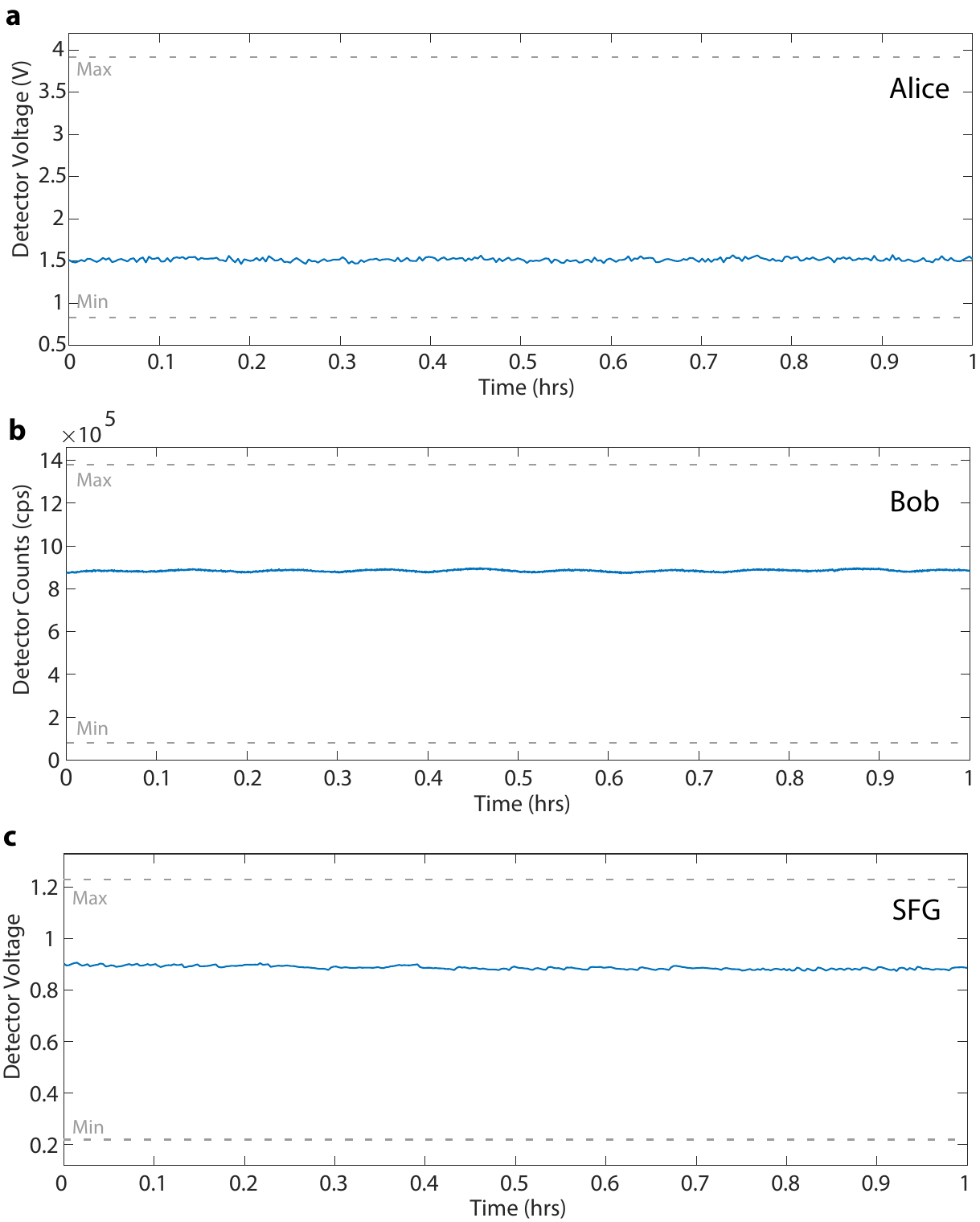}
		\caption{\textbf{Phase Stability.} \textbf{a-c} Transmission of continuous-wave light through each interferometer over time. The maximum and minimum of each fringe are marked with dashed lines. }
		\label{fig:stability}
	\end{center}
\end{figure*}

\section{Data analysis}

\subsection{Density matrix of the teleported state}
We denote the result obtained from projection measurements $\vert 0 \rangle \langle 0 \vert$,  $\vert 1 \rangle \langle 1 \vert$,  $\vert + \rangle \langle + \vert$,  $\vert - \rangle \langle - \vert$,  $\vert L \rangle \langle L \vert$,  $\vert R \rangle \langle R \vert$ as $N_0$, $N_1$, $N_+$, $N_-$, $N_L$, $N_R$, respectively ($\ket 0\equiv \ket e$ and $\ket 1\equiv \ket l$). The Stokes parameters of the time-bin qubit can be written as:

\begin{equation}
\begin{gathered}
	S_0 = N_0+N_1, \\
	S_1 = N_+-N_-, \\
	S_2 = N_L-N_R, \\
	S_3 = N_0-N_1.
\end{gathered}
\label{eq:S}
\end{equation}
The density matrix of the qubit can be obtained using the Stokes parameters,
\begin{equation}
\begin{gathered}
	\rho = \frac{1}{2} (\sigma_0 + \frac{S_1}{S_0}\sigma_1 + \frac{S_2}{S_0}\sigma_2 + \frac{S_3}{S_0}\sigma_3),
\end{gathered}
\label{eq:rho}
\end{equation}
where $\sigma_i$ ($i$ = 0,1,2,3) denote the Pauli matrices. 

Experimentally, the projection measurement result is obtained from the idler photon counts in the three time bins (bin width of 350 ps) after the unbalanced interferometer, denoted as $e$, $l$, and $ll$, conditioned on the detection of the SFG photon in the middle bin. Specifically, we used the following relation  
\begin{equation}
	\begin{gathered}
		N_0 = n_{0,1,e}+n_{0,2,e}, \\
		N_1 = n_{0,1,ll}+n_{0,2,ll}, \\
		N_- = n_{0,1,l}, \\
		N_+ = n_{0,2,l}, \\
		N_L = n_{\frac{\pi}{2},1,l}, \\
		N_R = n_{\frac{\pi}{2},2,l},
	\end{gathered}
	\label{eq:N_math}
\end{equation}
where $n_{i,j,k}$ denotes the counts collected for $\varphi_B = i$, at SNSPD $j$, from $k_{\mathrm{th}}$ time bin.

The density matrix directly constructed from the raw projection measurement data can be unphysical because of measurement noises. We used the maximum likelihood estimation algorithm \cite{altepeter2005photonic} to construct the physical density matrix of the teleported state. The constructed physical density matrices of the teleported states for cavity photon number $\left<n_a\right>=80$ and 8 are plotted in Fig. \ref{fig:densitymatrix}, while those for $\left<n_a\right>=0.8$ are shown in Fig. 4.

\begin{figure}[!htb]
\begin{center}
\includegraphics[width=\columnwidth]{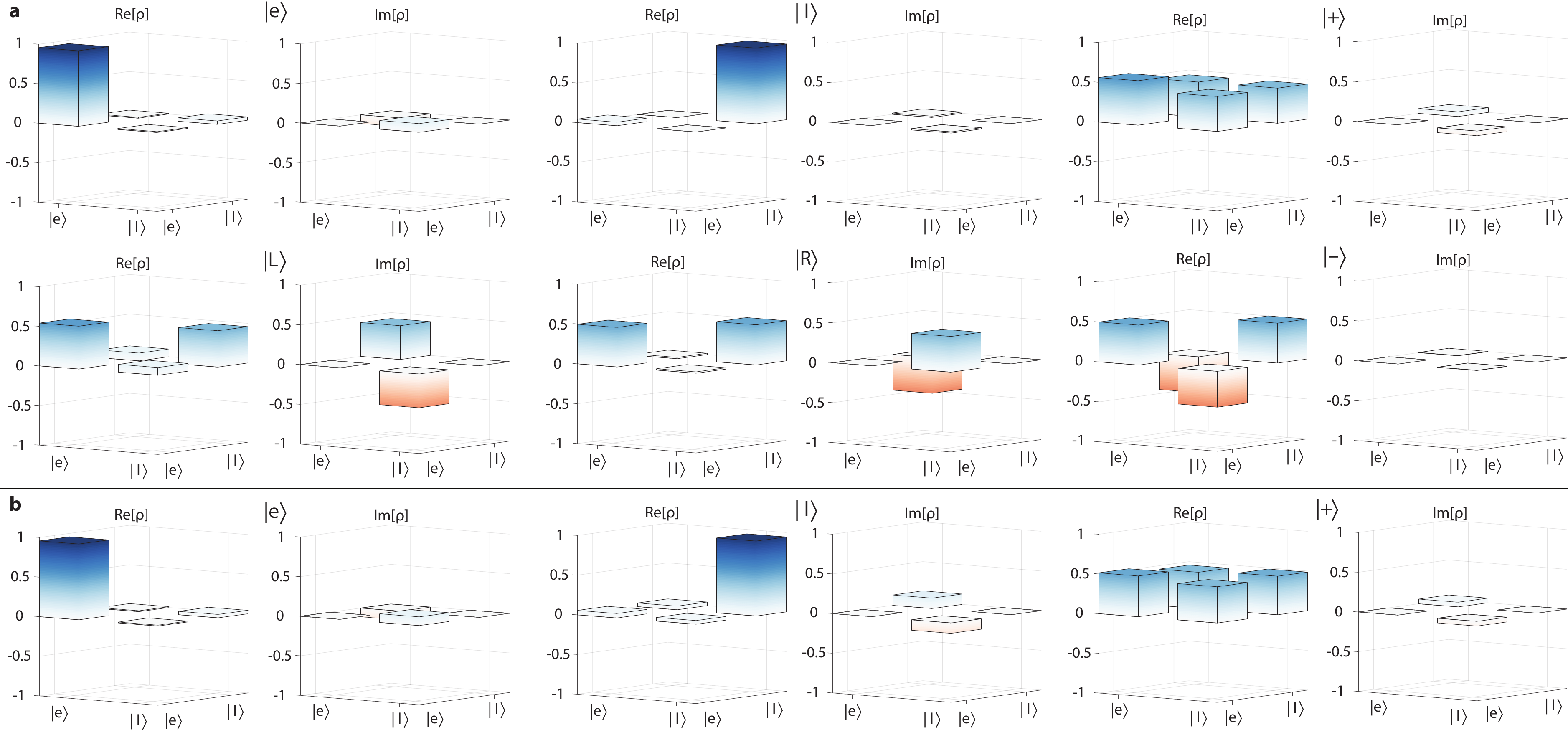}
\caption{\textbf{Density matrix.} Density matrix of the teleported states for  $\left<n_a\right>=80$ (\textbf{a}) and $\left<n_a\right>=8$ (\textbf{b}). }
\label{fig:densitymatrix}
\end{center}
\end{figure}

\subsection{Fidelity and purity of the teleported states}
Table \ref{tab:fidelity} summarizes the fidelity and purity of the teleported states $\ket e$, $\ket l$, and $\ket +$ for $\langle n_a\rangle=0.8,\ 8,\ 80$, which are calculated using an open source code \cite{kwiatcode}. 

\newcolumntype{L}[1]{>{\raggedright\let\newline\\\arraybackslash\hspace{0pt}}m{#1}}
\newcolumntype{C}[1]{>{\centering\let\newline\\\arraybackslash\hspace{0pt}}m{#1}}
\newcolumntype{R}[1]{>{\raggedleft\let\newline\\\arraybackslash\hspace{0pt}}m{#1}}
{\renewcommand{\arraystretch}{1.3}
\begin{table}[h!]
\centering
\caption{\textbf{Summary of fidelity and purity}}
\vspace{5pt}
\begin{tabularx}{ 0.78\textwidth} {  m{6cm} m{3.cm} m{3.5cm}  m{1cm}} 
    \hline
    \hline
    $\left<n_a\right>$ & State & Fidelity & Purity\\
    \hline
      0.8 & $\ket e$ & $95.5\pm1.6\%$ & 0.91 \\
        & $\ket l$ & $95.9\pm1.6\%$ & 0.93 \\
        & $\ket +$ & $94.8\pm2.3\%$ & 0.90 \\
       \hline
       8 & $\ket e$ & $95.7\pm1.4\%$ & 0.94 \\
        & $\ket l$ & $94.4\pm1.6\%$ & 0.93 \\
        & $\ket +$ & $95.7\pm1.9\%$ & 0.93 \\
        \hline
       80 & $\ket e$ & $94.9\pm1.3\%$ & 0.92 \\
        & $\ket l$ & $95.5\pm1.2\%$ & 0.92 \\
        & $\ket +$ & $93.9\pm2.0\%$ & 0.90 \\
        & $\ket -$ & $94.9\pm1.6\%$ & 0.90 \\
        & $\ket R$ & $94.7\pm2.1\%$ & 0.90 \\
        & $\ket L$ & $92.5\pm2.6\%$ & 0.88 \\
    \hline
    \hline
    \end{tabularx}
    \label{tab:fidelity}
\end{table}

\color{black}

\subsection{Teleportation fidelity error}
The teleportation fidelity can be calculated using $\mathcal{F} =\langle \psi_0 \vert \rho \vert \psi_0 \rangle$ for a desired state $\vert \psi_0 \rangle$ and the constructed density matrix $\rho$ of the teleported state (Eq. \ref{eq:rho}). The fidelity error is defined as
\begin{equation}
		\Delta \mathcal{F} =\langle \psi_0 \vert \Delta \rho \vert \psi_0 \rangle,
	\label{eq:dF}
\end{equation}
where the error of the density matrix of the teleported state, $\Delta \rho$, is calculated from Eq. \ref{eq:rho}, given variations of the measurement counts $\Delta N_i$. $\Delta N_i$ includes contributions from the shot noise and the variations of the interferometer phases, 
\begin{equation}
	(\Delta N_i)^2 = (\frac{\partial N_i}{\partial \theta} \Delta \theta)^2 + N_i.
\label{eq:Nerror}
\end{equation}

The shot noise contribution is calculated using Monte Carlo simulation, by applying Poissonian noise to experimental result $N_i$ and repeatedly calculate the density matrix and fidelity. Then the variation of fidelity is calculated using 100,000 simulations. 

The contribution from the phase variation is calculated as following. Note the phase variation only contributes to the superposition states.  For projected state $\psi_i(\theta)=\frac{1}{\sqrt{2}}(\vert 0 \rangle + \vert 1 \rangle e^{j\theta})$, where $\theta = 0,\pi,\pi/2,3\pi/2$ corresponds to $\vert + \rangle, \vert - \rangle, \vert L \rangle, \vert R \rangle$, respectively, according to the mechanism of projection measurement, we have:
\begin{equation}
	\begin{gathered}
		N_i = C \langle \psi_i(\theta)  \vert \rho \vert \psi_i(\theta)  \rangle,
	\end{gathered}
	\label{eq:N_i}
\end{equation}
where $C$ is a constant related to system efficiency and does not enter the final result of $\Delta F$.  From Eqs. \ref{eq:rho} and \ref{eq:N_i}, we have
\begin{equation}
\begin{gathered}
	\frac{\partial N_i}{\partial \theta} = \frac{C}{2} \sum_{k=0}^{3} \frac{S_k}{S_0} [(\frac{\partial}{\partial \theta} \langle \psi_i(\theta) \vert) \sigma_k \vert \psi_i(\theta) \rangle + \langle \psi_i(\theta) \vert\sigma_k \vert \frac{\partial}{\partial \theta} ( \psi_i(\theta) \rangle)].
\end{gathered}
\label{eq:part}
\end{equation}
The phase variation of Alice's and idler's glass MZI has a standard deviation of $\Delta\theta_{A, i}=1.1\times 10^{-3}\pi$ and that of the free-space interferometer has a standard deviation of $\Delta\theta_{\Sigma}=2.4\times 10^{-3}\pi$. Thus we estimate the joint phase error $\Delta \theta = \sqrt{\Delta\theta_{A}^2+\Delta\theta_{i}^2+\Delta\theta_{\Sigma}^2} = 2.9\times 10^{-3}\pi$. Using $\Delta \theta$ and Eq. \ref{eq:part}, we calculate the error of density matrix and fidelity caused by the phase variations.

\section{Estimated fidelity}

We can estimate the measured teleportation fidelity based on the probability of desired and undesired events, using \cite{valivarthi2016quantum}
\begin{equation}
		\mathcal{F} = \frac{P_\mathrm{D}}{P_\mathrm{D} + P_{\mathrm{UD}}},
	\label{eq:estF}
\end{equation}
where $P_\mathrm{D}$ and $P_\mathrm{UD}$ are probability of events generating desired and undesired coincidences, respectively. The estimation here assumes perfect interferometers.

To this end, we only consider events with $n_{si} \leq 2$ and Alice is in a coherent state. For example,  the event having one pair of signal and idler and starting with one pair from the SPDC source occurs with a probability:
\begin{equation}
	\begin{gathered}
		P(\alpha_A,1_s,1_i) = \frac{1}{4} |\alpha_A|^2 p_{si} p_{\mathrm{SFG}} t_{s}t_i \eta_i t_\Sigma \eta_\Sigma,
	\end{gathered}
	\label{eq:P111}
\end{equation}
where $|\alpha_A|^2$ is the cavity photon number of Alice, $p_{si}$ is the probability of generating one SPDC photon pair out of the SPDC source, $p_{\mathrm{SFG}}$ is the SFG probability, $t_{s}$, $t_i$, $t_\Sigma$ denote the transmission efficiency of signal, idler, and sum-frequency photons, and $\eta_i$ and $\eta_\Sigma$ denote the detector efficiencies.
Here $\frac{1}{4}$ counts the probability for these photons to be projected onto $\vert \Sigma^+ \rangle$. Similarly, we can write the possibility of other relevant events as:
\begin{equation}
	\begin{gathered}
		P(\alpha_A,2_s,1_i) = \frac{1}{4} |\alpha_A|^2 p_{si}^2 2p_{\mathrm{SFG}}  t_{s}^2 2  t_i\eta_i (1- t_i\eta_i) t_\Sigma \eta_\Sigma, \\
		P(\alpha_A,1_s,2_i) = \frac{1}{4} |\alpha_A|^2 p_{si}^2 p_{\mathrm{SFG}} 2 t_{s} (1-t_{s}) (t_i\eta_i)^2 t_\Sigma \eta_\Sigma, \\
		P(\alpha_A,2_s,2_i) = \frac{1}{4} |\alpha_A|^2 p_{si}^2 2p_{\mathrm{SFG}} t_{s}^2 (t_i\eta_i)^2 t_\Sigma \eta_\Sigma, \\
	  P(\alpha_A,1_s',1_i') = \frac{1}{4} |\alpha_A|^2 p_{si}^2 p_{\mathrm{SFG}} 2 t_{s} (1-t_{s}) 2  t_i\eta_i (1- t_i\eta_i) t_\Sigma \eta_\Sigma,
	\end{gathered}
	\label{eq:Pother}
\end{equation}
where the $(1_s',1_i')$ in the last term is derived from $(2_s,2_i)$ due to photon loss. 

The desired probabilities for $\pm, L, R$ states are
\be
P_\mathrm{D} = P(\alpha_A,1_s,1_i) +0.5P(\alpha_A,2_s,1_i)  + P(\alpha_A,1_s,2_i) + P(\alpha_A,2_s,2_i)+ 0.5P(\alpha_A,1_s',1_i').
\ee
The reason of the 0.5 factor in the second term is that, for a successful teleportation, the signal photon converted to a SFG photon should be entangled with the transmitted idler photon. Similarly, for the 0.5 factor of the last term, the transmitted signal and idler photons have to be one of the entangled pairs out of the source. 
The estimated fidelity thus can be expressed in terms of system efficiency:
\bqa
\nonumber\mathcal{F}_{\pm, L, R} &=& \frac{P_D}{P_D + P_{UD}} \\\nonumber
&=& \frac{P(\alpha_A,1_s,1_i) +0.5P(\alpha_A,2_s,1_i)  + P(\alpha_A,1_s,2_i) + P(\alpha_A,2_s,2_i)+ 0.5P(\alpha_A,1_s',1_i')}{P(\alpha_A,1_s,1_i) +P(\alpha_A,2_s,1_i)  + P(\alpha_A,1_s,2_i) + P(\alpha_A,2_s,2_i)+ P(\alpha_A,1_s',1_i')}\\\nonumber
&=& 1-\frac{2 p_{si} t_{s} (1-t_i\eta_i) +2p_{si}  (1-t_{s})  (1- t_i\eta_i)}{1+4 p_{si} t_{s} (1-t_i\eta_i) + 2 p_{si} (1-t_{s}) t_i\eta_i + 2 p_{si} t_{s} t_i\eta_i+4 p_{si}  (1-t_{s})  (1- t_i\eta_i)}\\
&=&\frac{1+2p_{si}}{1+2p_{si}(2-t_i\eta_i )}.
\eqa

The desired probability for $e, l$ states are
\be
P_\mathrm{D} = P(\alpha_A,1_s,1_i) +0.75P(\alpha_A,2_s,1_i)  + P(\alpha_A,1_s,2_i) + P(\alpha_A,2_s,2_i)+ 0.75P(\alpha_A,1_s',1_i').
\ee
The reason of the 0.75 factor is that besides the 50\% probability that signal and idler belong to the same entangled photon pair, there is additional $50\%\times50\%=25\%$ probability that the signal and idler end in the same time bin even though they do not belong to the same entangled photon pair.
The estimated fidelity thus can be expressed as:
\bqa
\nonumber\mathcal{F}_{e,l} &=& \frac{P_D}{P_D + P_{UD}} \\\nonumber
&=& \frac{P(\alpha_A,1_s,1_i) +0.75P(\alpha_A,2_s,1_i)  + P(\alpha_A,1_s,2_i) + P(\alpha_A,2_s,2_i)+ 0.75P(\alpha_A,1_s',1_i')}{P(\alpha_A,1_s,1_i) +P(\alpha_A,2_s,1_i)  + P(\alpha_A,1_s,2_i) + P(\alpha_A,2_s,2_i)+ P(\alpha_A,1_s',1_i')}\\\nonumber
&=& 1-\frac{ p_{si} t_{s} (1-t_i\eta_i) +p_{si}  (1-t_{s})  (1- t_i\eta_i)}{1+4 p_{si} t_{s} (1-t_i\eta_i) + 2 p_{si} (1-t_{s}) t_i\eta_i + 2 p_{si} t_{s} t_i\eta_i+4 p_{si}  (1-t_{s})  (1- t_i\eta_i)}\\
&=&\frac{1+p_{si}(3-t_i\eta_i )}{1+2p_{si}(2-t_i\eta_i )}.
\eqa

Based on the measured system efficiency presented in Table \ref{tab:loss} and $p_{si}=0.003$ (in a bandwidth matched to the Alice's pulse in the cavity), the estimated fidelities of all six states are $>99\%$. This near-perfect fidelity explains our measured high fidelity well. The remaining difference can be attributed to imperfections in the interferometers.

\section{Complete Bell state analyzer}

\begin{figure*}[!htb]
	\begin{center}
		\includegraphics[width=0.7\columnwidth]{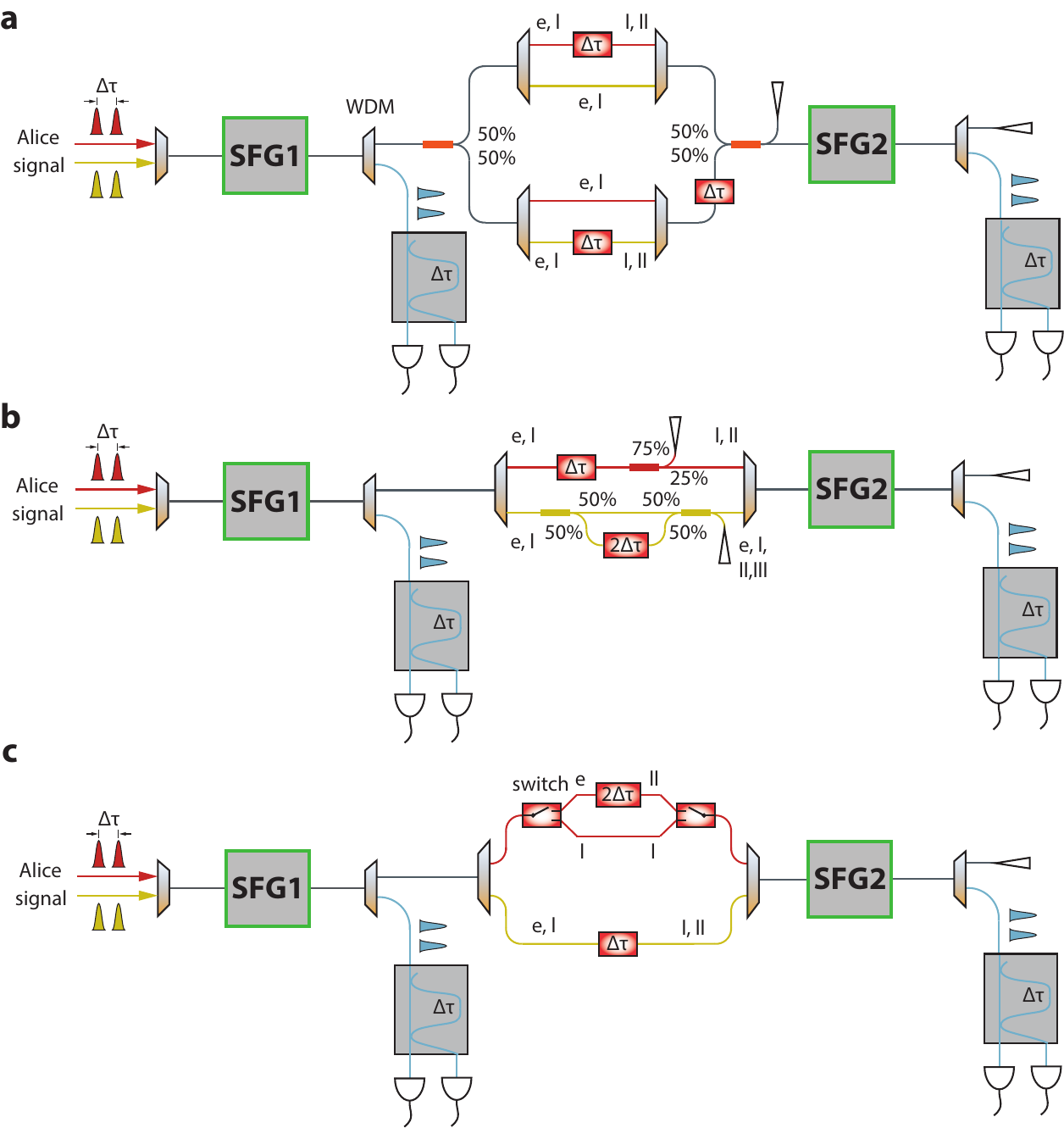}
		\caption{\textbf{Complete Bell state analyzer.} \textbf{a} and \textbf{b}. Two ``lossy''schemes using beamplitters. \textbf{c}  A ``lossless'' scheme using optical swithces. }
		\label{fig:completeBSA}
	\end{center}
\end{figure*}

A complete Bell state analyzer can be implemented with two nonlinear SFG elements. To show this, suppose Alice's photon is prepared in a superposition state of the early and late time bins, $\ket{\psi_A}=\alpha \ket e_\mathrm{A}+\beta \ket l_\mathrm{A}$ and the entangled photon source generates a photon pair in the Bell state $\ket{\Phi^+}_{12}=\frac{1}{\sqrt{2}}(\ket e_1\ket e_2+\ket l_1\ket l_2)$, where the subscripts 1 and 2 denote signal and idler photons, respectively. The joint state of the three photons can be expressed as 
\bqa
\ket{\psi_A}\otimes\ket{\Phi^+}
&=&\frac{1}{2}\big(\ket{\Phi^+}_{A1}(\alpha\ket e_2+\beta\ket l_2)+\ket{\Phi^-}_{A1}(\alpha\ket e_2-\beta\ket l_2)\\\nonumber
&&+\ket{\Psi^+}_{A1}(\alpha\ket l_2+\beta\ket e_2)+\ket{\Psi^+}_{A1}(\alpha\ket l_2-\beta\ket e_2) \big),
\eqa
where $\ket{\Phi^\pm}_{A1}=\frac{1}{\sqrt{2}}(\ket e_A\ket e_1\pm\ket l_A\ket l_1)$ and  $\ket{\Psi^\pm}_{A1}=\frac{1}{\sqrt{2}}(\ket e_A\ket l_1\pm\ket l_A\ket e_1)$ are the four Bell states.  The goal is to convert the four Bell states to four distinct SFG single-photon states, and thus two SFG elements are required. One SFG element will interact $e(l)$ with $e(l)$ and convert $\ket{\Phi^+}_{A1}$ to two orthogonal SFG states, while the other SFG element will only interact $e$ with $l$ and convert $\ket{\Psi^+}_{A1}$ to two orthogonal SFG states in different spatial modes. After the complete Bell state analyzer, the joint state of the initial three photons becomes 
\bqa\label{sfgteleport}
\ket{\psi_A}\otimes\ket{\Phi^+}_{12}&\xrightarrow{\text{SFG}}&\frac{1}{2}\big(\ket{\Sigma_1^+} (\alpha\ket e_2+\beta \ket l_2)  +\ket{\Sigma_1^-}(\alpha\ket e_2-\beta \ket l_2) \\\nonumber 
&& +\ket{\Sigma_2^+} (\alpha\ket l_2+\beta\ket e_2)  +\ket{\Sigma_2^-}(\alpha\ket l_2-\beta\ket e_2)  \big),
\eqa
where $\ket{\Sigma_{1(2)}^{\pm}}=\frac{1}{\sqrt{2}}( \ket e_{\Sigma_{1(2)}}\pm \ket l_{\Sigma_{1(2)}})$ are the orthogonal SFG photon states generated by the first and second nonlinear elements. The four SFG states distinguish the four Bell states completely. By measuring the SFG photon in one of the states, the idler photon is projected to a single-photon state, which differs from the original Alice's photon state up to a single-qubit rotation. 

Two schemes of the complete Bell state analyzer are illustrated in Fig. \ref{fig:completeBSA}. For the first scheme (Fig. \ref{fig:completeBSA}a), the first nonlinear element induces interaction between $ee$ or $ll$ time bins to generate $e$ and $l$ SFG photons, whose amplitudes interfere at an unbalanced MZI for projection measurement. Before the second nonlinear cavity, Alice and signal photons pass through two sets of wavelength-selective (via WDMs) delay lines to delay Alice and signal and temporally align $\ket e_A\ket l_1$ and $\ket l_A\ket e_1$, respectively. $\ket e_A\ket e_1$ and $\ket l_A\ket l_1$ will be temporally mismatched after the delay lines. The second nonlinear element then induces interaction only between the $e$ and $l$ time bins, and the generated $e$ and $l$ SFG photons (because of the relative delay between the two sets of delay lines) interfere at an unbalanced MZI for projection measurement. In the second scheme (Fig. \ref{fig:completeBSA}b), after the same first SFG element, wavelength-selective delay lines are introduced to convert Alice photon bins $e\rightarrow l$, $l\rightarrow ll$ and signal photon bins $e\rightarrow (e,ll)$,  $l\rightarrow (l,lll)$. Thus, only $\ket e_A\ket l_1$ and $\ket l_A\ket e_1$ become temporally aligned and can generate SFG photons in the second nonlinear element. Using optical switches in replacement of beamsplitters, the nonlinear Bell state analyzer can be made ``lossless'' (aside from the finite SFG efficiency), as shown in Fig. \ref{fig:completeBSA}c as an example.